\documentclass[10pt,sigconf,letterpaper]{acmart}
\usepackage[english]{babel}
\usepackage{blindtext}
\usepackage{subfigure}
\usepackage{listings}
\usepackage{footnote}
\usepackage{tablefootnote}
\usepackage{multirow}
\usepackage{graphicx}
\usepackage{makecell}
\usepackage{comment}
\usepackage{totpages}

\usepackage{algorithm}
\usepackage{algpseudocode}
\floatname{algorithm}{Pseudocode}

 \renewcommand\footnotetextcopyrightpermission[1]{} 

\newcommand{\name}{\'Onoma}
\usepackage{xassoccnt}

\NewTotalDocumentCounter{totalpages}
\DeclareAssociatedCounters{page}{totalpages}
\usepackage{blindtext}

\usepackage{forloop}
\newcounter{loopcntr}

\begin{document}
\title{Reclaiming Privacy and Performance over Centralized DNS}

\author{Rashna Kumar}
\affiliation{%
	\institution{Northwestern University}
}
\email{rashnakumar2024@u.northwestern.edu}

\author{Fabi\'an E. Bustamante}
\affiliation{%
	\institution{Northwestern University}
}
\email{fabianb@cs.northwestern.edu}



\begin{abstract}

The Domain Name System (DNS) is both a key determinant of users’ quality of experience (QoE) and privy to their tastes, preferences, and even the devices they own. Growing concern about user privacy and QoE has brought a number of alternative DNS services, from public DNS to encrypted and Oblivious DNS. While offering valuable features, these DNS variants are operated by a handful of providers, reinforcing a trend towards centralization that has raised concerns about privacy, competition, resilience and Web QoE. 

The goal of this work is to let users take advantage of third-party DNS services, without sacrificing privacy or performance. We follow Wheeler's advice, adding another level of indirection with an end-system DNS resolver, {\it \name}, that improves privacy, avoiding DNS-based user-reidentification by inserting and sharding requests across resolvers, and improves performance by running resolution races among resolvers and reinstating the client-resolver proximity assumption content delivery networks rely on. As our evaluation shows, while there may not be an ideal service for \textit{all} clients in \textit{all} places, \name~dynamically finds the \textit{best} service for \textit{any} given location. 

\end{abstract}
\maketitle

\section{Introduction}
\label{sec:introduction}

Domain Name System (DNS) is a key component of the Internet. Originally designed to translate domain names to
IP addresses~\cite{mockapetris:rfc1034}, it has evolved into a complex infrastructure providing platform for a
wide range of applications.

Today DNS is a key determinant, directly and indirectly, of users' quality of experience (QoE) and privy to their 
tastes, preferences, and even the devices they own. It directly determines user performance as, for instance, 
accessing any website requires tens of DNS resolutions~\cite{butkiweicz:complexity, bottger:dohqoe,bozkurt:slow}. %
Indirectly, a user's specific DNS resolver determines their QoE as many content delivery networks 
(CDNs) continue to rely on DNS for replica selection. %
As we show in Sec.~\ref{sec:background}, despite some adoption of anycast~\cite{calder:anycastcdn,alzoubi:ayncastcdn} 
and EDNS Client Subnet (ECS)~\cite{chen:akamai}, and the promise of CDN-ISP  collaborations~\cite{pujol:steering,poese:padis,xie:p4p,alimi:alto}, 
many CDNs continue to rely on the assumption that the location of a client's resolver provides a good proxy for the client's own location~\cite{hounsel:ddns,colder:edns-adoption}.

Beyond its critical role in the user's experience, clear text DNS requests can have a significant impact on privacy and security 
and, in certain parts of the world, on human rights~\cite{dnscensor:venezuela, dnscensor:iran, dnscensor:ccr12}. %
The set of DNS requests issued by a user reveals much about their tastes, preferences, or the devices they own and how they used them~\cite{bortzmeyer:dnsprivacy,gchq,morecowbell,ghuston:dnsprivacy}, even when using VPN or Tor~\cite{greschbach:dnstor}. %
Prior work~\cite{olejnik:browsinghist,bird:browsinghist} shows that browsing profiles, a subset of users' DNS request streams, 
are both highly distinctive (e.g., 98-99\% of browsing profiles are unique) and stable (e.g., re-identifiability rate as high 
as ~80\%), suggesting the viability of browser profiles as user identifiers. 

Concerns about user privacy and QoE have served as motivation for a number of alternative DNS techniques and third-party services, 
from public DNS to encrypted and Oblivious DNS~\cite{schmitt:odns,odns:cloudflare,kinnear:adaptivedns,schmitt:odns}. %
Google Public DNS was announced in late 2009 promising better performance and higher reliability~\cite{google:publicdns}, 
while DNS over HTTPS (DoH), DNS over TLS (DoT), DNSCrypt, and Oblivious DNS are some of the latest proposals to make DNS more secure~\cite{liu:doe,odns:cloudflare,dnscrypt:www}.

While offering some valuable features, these DNS variants are operated by a handful of providers such as Google, Cloudflare or 
IBM, strengthening a concerning trend toward DNS centralization and its implications~\cite{ietf:consolidation,huston:centralization, wang:consolidation}.
First, the use of encrypted DNS, while avoiding man-in-the-middle attacks, does not necessarily improve client's privacy as the DNS provider has access to the highly-unique, unencrypted request streams of millions of clients and is a clear target for a court order to release data in bulk~\cite{valentino:subpoenas}.
Second, while centralized DNS could offer better DNS response times, it may result in worst Web QoE as it breaks CDNs' assumption 
about the proximity between users and their resolvers~\cite{otto2012content}. Our evaluation (\S\ref{sec:system_design}) shows that, despite the large-scale deployment of many third-party DNS services, the proximity to users of their resolvers varies widely across locales~\cite{colder:edns-adoption}. While EDNS IP Subnet extension is meant to address this problem, prior work~\cite{colder:edns-adoption} shows that adoption of ECS across the Internet is minuscule and even large providers, such as Cloudflare, have stopped supporting it due to privacy concerns~\cite{cloudflare:ecs}. \footnote{The Cloudflare's CEO commented that his company is instead exploring an alternative to the EDNS IP Subnet extension for user geolocation~\cite{cloudflare:ecs}.} 
Last, part of the Internet's inherent resilience lies in its diversity; centralization leads to a clustering of multiple risks, from technical to economic ones~\cite{huston:centralization}, while increasing the potential impact of any single failure~\cite{cloudflare:down}.


\textit{The goal of our work is to let users take advantage of third-party DNS services without sacrificing privacy or performance. Our approach follows Wheeler's advice, adding another level of indirection with an end-system DNS resolver.}\footnote{``All problems in computer science can be solved by another level of indirection'', David Wheeler}
We present {\it \name}, an end-system resolver that:
\begin{enumerate}
\item enhances privacy by sharding users' requests across \textit{multiple} resolvers~\cite{hoang:sharding,hounsel:ddns,jari:sharding} (\S\ref{subsubsection:domainsharding}),
\item reduces reidentifiablitly by adaptively inserting popular requests on a user DNS request stream based on the uniqueness of a requested domain (\S\ref{subsubsection:insertion}),
\item improves DNS performance, independently of a user's location and specific DNS or CDN services' deployments, by running resolution races~\cite{vulimiri:racing} among resolvers (\S\ref{subsubsection:racing}) and 
\item addresses the QoE impact of client-LDNS mismatch problem~\cite{otto2012content} (\S\ref{subsubsection:directresolution}). 
\end{enumerate}

We evaluate \name~across geographic locales and with different DNS services, content providers, and CDNs and a range of settings for sharding, insertion and racing. %
Our evaluation results show that \name~can take advantage of third-party DNS while avoiding user re-identification -- 100\% of the users can not be re-identified (are k-anonymous) with low insertion and sharding requests across 8 resolvers (\S\ref{subsection:privacy}). Our results also demonstrate that, across a number of DNS resolvers and CDNs, \name~can improve average resolution time by 38.8\% by racing just two resolvers, and can yield up to ~10\% shorter Time-To-First-Byte(\S\ref{subsec:performance}). More importantly, the average performance of \name~is better or equal to the best service for most CDN and individual DNS at any locale (\S\ref{sec:evaluation}).
While there may not be an ideal service for all clients in \textit{all} places, \name~dynamically selects the \textit{best} service for \textit{any} given location. 




\section{Background}
\label{sec:background}

In the following paragraphs, we provide background on DNS, the relation between DNS and CDNs and recent trends in the DNS market.

The process of DNS resolution involves at least a stub resolver that asks questions from a recursive resolver which, in turn, finds the 
answer by asking one or more authoritative DNS servers. Traditionally, the user's recursive resolver has been provided by the user's ISP 
and found, in networking terms, generally close to their users. 

ISP DNS recursive resolvers are, however, not always the best in terms of resolution time~\cite{ager:dns,rula:behind}, and slower DNS 
resolutions can significantly impact users' QoE. Besides resolution performance, there are several other potential drawbacks of using 
an ISP provided resolver in terms of reliability, privacy and censorship~\cite{comcast:outage,ripe:censorship,liu:ispPrivacy}. 

Partially in response to these issues, a third-party ecosystem has evolved around DNS over the years. Services such as OpenDNS, UltraDNS 
and Google Public DNS~\cite{google:publicdns} offer a number of valuable features, from load balancing, nonce prepending and shared caching 
for performance, to phishing and botnet detection for security. More recently, some of these same providers have begun to offer DNS over 
HTTPs (DoH), DNS over TLS (DoT) and DNSCrypt~\cite{dnscrypt:www} using encryption between the DNS stub and the provided recursive resolver 
to prevent on-path eavesdropping, spoofing, and blocking~\cite{cloudflaredoh,quad9doh}. These DNS services typically host servers on multiple 
sites around the globe and use anycast to guide DNS clients to the nearest resolver. Note, of course, that the resolution times experienced 
by any particular client will depend in part on their location and proximity to these ``nearest'' resolvers for their chosen DNS service.

The proximity, or lack thereof, between clients and their DNS resolvers can impact not just resolution times. This observation has been at
the core of CDNs mapping systems, sometimes referred to as \textit{redirection}, the process by which the CDN decides the replica server 
where to redirect a client's request for content to improve user-perceived performance and balance the load across servers~\cite{hao:sec18}. 

The proximity, or lack thereof, between clients and their DNS resolvers can impact not just resolution times, depending on the CDN redirection 
method. While most of the time DNS-based redirection performs well, the ``client-LDNS mismatch'' hurts performance 
by directing users to distant servers. EDNS Client Subnet (ECS) aims to address this problem by allowing a client's resolver to forward a 
portion of the client’s IP address to an authoritative DNS server. While a number of large-scale DNS services and some CDNs have adopted ECS, 
older et al.~\cite{colder:edns-adoption} shows that adoption of ECS across the Internet is minuscule and even large providers, such as Cloudflare, 
have stopped supporting it due to privacy concerns~\cite{cloudflare:ecs}. 

To illustrate the potential impact of client-LDNS mismatch, we select top50 Alexa sites\footnote{We discuss the use of Alexa regional rankings in 
Sec.\ref{sec:limitation}.}  in the US and resolve the resources of these sites, hosted on the two most popular CDNs -- Akamai and Amazon CloudFront, 
using the Ripe Atlas DNS API. We use probes in the US and the RIPE probes' resolvers as local resolvers, and use distant regional resolvers in 
Argentina and India. We measure the median round-trip time (RTT) of five runs to the replicas assigned using the local and distant resolvers. If 
a CDN does not rely on DNS for redirection, one would expect no RTT differences between the replicas assigned when using the different resolvers. 
If, on the other hand, a CDN does rely on DNS-based redirection, there should be larger RTTs to the replicas assigned with increasingly distant 
resolvers, even when the client location does not change.


\begin{figure}
	\centering
	\subfigure[Akamai CDN\label{fig:Akamai_DNS_localization}]{\includegraphics[width=0.8\columnwidth]{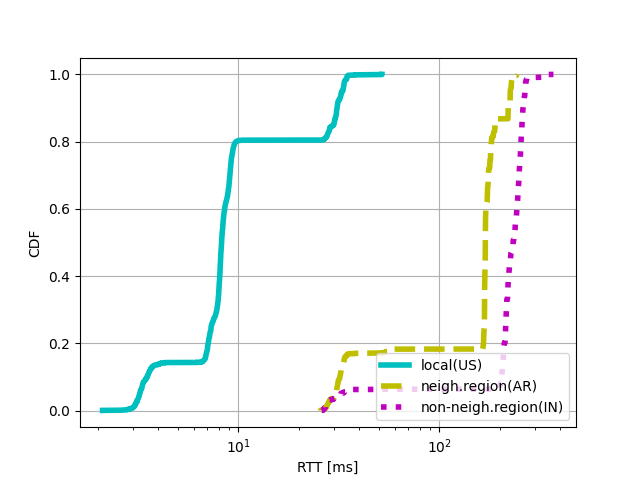}}	
	\caption{\textit{Client-LDNS mismatch problem and CDN redirection.} Latency to assigned replica servers for Akamai and Amazon CloudFront for resources in top50 US websites. The client in the US is using local DNS resolvers (cyan) and distant resolvers in Argentina and India.}
	\label{fig:DNS_localization}
\end{figure}

Figure~\ref{fig:DNS_localization} clearly illustrates the client-LDNS mismatch problem with the two most popular CDNs (full set of graphs in \S\ref{sec:appendix}). The RTTs to replicas assigned when using a local resolver are consistently lower - up to an order of magnitude lower at the median than those to replicas assigned when using \textit{any} of the distant resolvers (e.g., 9ms for the local resolver and 160ms with a resolver in India in the case of Akamai), and the overall distributions are ordered matching the relative geographic distances of the replicas.

\section{\name's Motivations and Goals}
\label{sec:goals}

Concerns about user privacy and QoE have served as motivation for a number of alternative DNS techniques and third-party services~\cite{schmitt:odns,odns:cloudflare,kinnear:adaptivedns,google:publicdns,odns:cloudflare,dnscrypt:www}. While 
offering some valuable features, these DNS variants are operated by a handful of providers such as Google, Cloudflare or 
IBM, strengthening a concerning trend toward DNS centralization and its implications~\cite{ietf:consolidation,huston:centralization, wang:consolidation, ietf:consolidation}. We briefly discuss these implications, as they motivate the design of \name, 
and present its derived goals. 

For starters, DNS-over-Encryption services are not privacy panacea. While they protect user data from ISPs and 
man-in-the-middle attacks, they expose all information to the recursive DNS providers.  Prior studies~\cite{olejnik:browsinghist, bird:browsinghist} have analyzed the uniqueness of users' web browsing histories, a subset of their DNS request streams, 
and shown that users' browsing profiles are both highly distinctive and stable (and thus a viable tracking vector). 
Centralized DNS is also an easier target for court order to release data in bulk~\cite{valentino:subpoenas} and no 
privacy agreement policy by a DNS provider can escape that. Recent work on Oblivious DNS addresses some of these concerns 
by preventing queried domains from being associated with the client identity~\cite{schmitt:odns,odns:cloudflare}. However, Oblivious DNS also increases the cost of DNS resolutions and contributes to the client-LDNS mismatch problem impacting users' web QoE.

In addition, consolidation may also result in lower infrastructure resilience and increases the risk of a captive market~\cite{ietf:consolidation}. Using measurements from 100,000 users of the OONI app, Radu et al.~\cite{radu:centralization} 
show that Google and Cloudflare control 49.7\% of the resolver market. Part of the Internet's inherent resilience lies 
in its diversity (i.e., different networks, peerings, code base and servers), and  centralization leads to a clustering 
of multiple such risks, at once increasing the probability of failures \textit{and} the potential impact of any single failure~\cite{cloudflare:down,abhishta:dyn}. 


Finally, while many of these infrastructures are pervasive~\cite{calder:mapping}, for specific locales a particular 
third-party DNS service may not have nearby servers to offer~\cite{otto2012content}, negatively impacting both DNS 
resolution times and web QoE for users of the many CDNs that continue to rely on DNS for replica selection. Poor 
user-mappings can lead to increased delays and worsening QoE, lower engagement~\cite{kirshnan:quasi} and revenue 
loss~\cite{whitenton:needspeed}.

\textbf{\name's Goals.} The emergence of a handful of third-party recursive DNS services is a natural result of a maturing Internet market and the benefits of economies of scale. \name's design aims to let users take advantage of third-party DNS services while avoiding the potential costs. Specifically, \name~aims to improve privacy by avoiding re-identification of users based on their DNS request streams, and achieve performance comparable to that of the best performing services in the user's locale, independently of any deployment of any specific DNS service, while avoiding reliance on any single DNS service. In addition, \name{} should be an easy-to-install, readily-deployable solution that bypasses the need for agreements and coordination between providers (CDNs, DNS or ISPs).

\section{System Design}
\label{sec:system_design}


In the following paragraphs, we present the key features of \name's design, 
experimentally motivating and illustrating the benefits of each of its features. We 
close the section with an overview of \name's system design. 

\subsection{Privacy and Third-party DNS}
 
A recursive DNS resolver is privy to much information about a user's taste, preferences and habits. 
Prior studies~\cite{olejnik:browsinghist,bird:browsinghist} have analyzed the uniqueness of users' 
web browsing histories. Olejnik et al.~\cite{olejnik:browsinghist}, among other findings, show that 
out of the 382,269 users that completed their popular site test, 94\% had unique browsing histories. 
They found that with just the 50 most popular sites in their users' histories -- a subset of their 
DNS request streams -- they were able to get a very similar distribution of distinctive histories 
compared to complete knowledge of the site list. The replication study by Bird et al.~\cite{bird:browsinghist} 
confirm this and show that such histories are also highly stable and thus a viable tracking vector. 

Beyond having access to detailed DNS request traces of million of users~\cite{google:publicdns}, 
large third-party services are also an easier target for court order to release data in 
bulk~\cite{valentino:subpoenas} and no privacy agreement policy can escape that. 

\name{} improves privacy by avoiding DNS-based user identification through $(i)$ domain specific sharding~\cite{hoang:sharding,hounsel:ddns,jari:sharding} and $(ii)$ DNS request insertion.

\subsection{Evaluating Privacy Gains}
\label{subsection:privacy} 

Since DNS resolvers are, by definition of their functionality, privy to the unencrypted version of the 
requested resources, the best privacy a user could expect is in terms of k-anonymity. A dataset is said to 
be k-anonymous if the information for each person contained in the release cannot be distinguished from at 
least $k - 1$ individuals whose information also appear in the release. To measure the k-anonymity value, 
we group users with the same profile (computed using jaccard distance between the profiles) in the same 
cluster. Users in clusters of size 1 (k-value = 1), can be re-identified on the basis of unique profiles. 

We evaluate the privacy gains of the different techniques used by \name~in terms of their impact on k-anonymity. 
We use for this two weeks of users' DNS logs, separated by one week, collected (using CLI~\cite{oracle:cli} and 
Splunk~\cite{splunk:www}) and made available (anonymized) by the Information Technology Center (IT) team at our 
institution. This two-week log correspond to 85,967 users. For every user, the DNS request logs consist of a 
browsing history, over a window of time, including all urls resolved by the user during that time. 

From the DNS request logs of every user, we construct a boolean vector of distinct domains, thereby 
removing all frequency measures. This boolean vector comprises the union of the 50 most popular domains 
visited by each user in our set and ranks the domains based on their popularity across all users. The 
length of our data collection window, and the format and size of the vector of popular domains we used, 
follows Olejnik et al.~\cite{olejnik:browsinghist} and Bird et al.~\cite{bird:browsinghist}'s approaches.

\paragraph{Ethics.} Since DNS requests are by their nature sensitive, some ethical questions arise that 
we address here. 

To preserve the privacy of our users, the data we receive from the IT team consists of the sites visited by 
each user stored against random non-persistent ids so our data does not give us access to any personal 
identifiers of users. These ids are one-way hashes and can not be mapped back to the users. They are, however, 
unique for each user allowing us to check if we can correctly re-identify users (anonymized) based on their 
DNS request logs. Since we are not able to identify individual users, we cannot 
associate the specific queries of any sensitive nature with any particular user. At most we can only identify 
the queries that are being made.

In addition, we submitted our methodology to our institution’s ethical review board and were given a 
determination of ``not human research''.

\subsubsection{Domain-specific Sharding}
\label{subsubsection:domainsharding} 

Previous work has suggested \textit{sharding} -- scattering DNS requests, randomly, across different resolvers -- as a way to address centralization concerns and to improve privacy~\cite{hoang:sharding,hounsel:ddns,jari:sharding}. Sharding prevents any single resolver from acquiring the complete user's request stream. However, if all DNS requests were distributed randomly across the available DNS servers, after a sufficiently long period of time, each DNS providers will have been queried for most DNS questions. Thus, a basic random distribution of request across resolvers makes each DNS service provider able to obtain a full view of the client's request stream and construct their profile. 

\name{} avoids this by using domain-specific sharding~\cite{arko:centraliseddns}. After selecting a DNS provider for a query, \name{} stores this mapping in a domain-to-DNS map. All of subsequent queries of a user under the same SLD will be directed to the same DNS server. This form of sharding~\cite{jari:sharding} limits the view of DNS service providers to a subset of the users' queries, preventing any provider from constructing a full user profile.

To evaluate the privacy benefits of domain specific sharding, we select users, from the two-week data set, with DNS request streams including more than 50 unique SLDs.\footnote{We ran our analysis over a window of different days in the two-week period and obtained similar results.} We apply the domain specific sharding algorithm on the request streams of each user and distribute the requests across a set of resolvers, $R$, which we referred to as the sharding value.

Each resolver stores a bit vector for each user (a union of the 50 most popular domains visited by all users, initially set to zero) and sets the bit to one for the user if the corresponding domain is found in the user's request stream. Our privacy goal is to challenge re-identification, ensuring that the vector maps at each resolver for different users are similar for several ($k > 1$) users.

\begin{figure}[h]
	\centering
	\includegraphics[width=0.8\columnwidth]{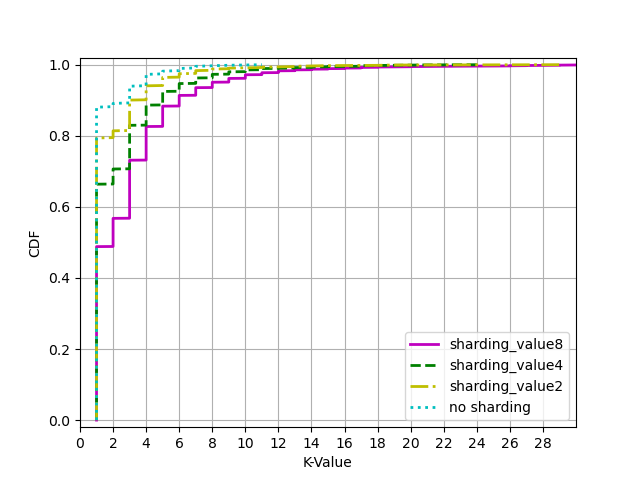}
	\caption{K-anonymity values of the set of users with different sharding values.}
	\label{fig:sharding}
\end{figure}

Figure~\ref{fig:sharding} shows the cumulative distribution function (CDF) of k-anonymity, averaged across resolvers, for different sharding values. \textit{No sharding} provides the baseline and has the lowest level of k-anonymity with k-values of 1, i.e., unique profiles, for over 88\% of users. This is consistent with previous findings on the uniqueness of browsing histories~\cite{olejnik:browsinghist,bird:browsinghist}. A sharding value of 2, that is when a user's request stream is sharded across two resolvers, drops the set of users that can be re-identify to 79\%; the maximum sharding value of 8 in this analyzes reduces this to 49\%. A concern with high sharding values is that while relying on more resolvers may improve improve k-anonymity, it also increases the chance of finding an under-performing resolver which can impact the user's average resolution time. We illustrate this and discuss \name~response in Sec.~\ref{subsec:performance}.

\subsubsection{Popular Request Insertion}
\label{subsubsection:insertion} 

Domain specific sharding, while clearly useful, has limited impact on k-anonymity. \name~further improves the k-anonymity of users by adaptively inserting popular requests on a users' DNS request stream based on its uniqueness. The more distinctive the request streams,
the higher the fraction of popular DNS request streams needed to make it similar (higher $k$) to other users' request streams.\footnote{\name~lets users to direct truly distinctive or otherwise problematic domains to an Oblivious DNS~\cite{odns:cloudflare,schmitt:odns}, thus limiting its performance impact.} 

Our current implementation inserts a number, an ``insertion factor'', of popular domains randomly sampled from the top50 most popular sites,\footnote{We exclude a configurable blacklist of domains.} for every \textit{unique} domain in the users' request stream. 
We consider a domain as \textit{unique} if it is not part of the top500 most popular sites. The approach is clearly independent of both the specific 
list and the subsets (top50 and top500) we use.

To analyze the privacy benefits of popular request insertion, we follow the same approach as in our evaluation of sharding, and apply our insertion algorithm with different values of the insertion factor $f$. As before, each recursive resolver stores a bit vector per user, recomputed for each value of $f$ as insertion causes an addition of requests to the users' request streams. Each resolver, as before, sets the bit to one for a given user if it sees the corresponding domain in the user's request stream. 

\begin{figure}[ht]
	\centering
	\includegraphics[width=0.8\columnwidth]{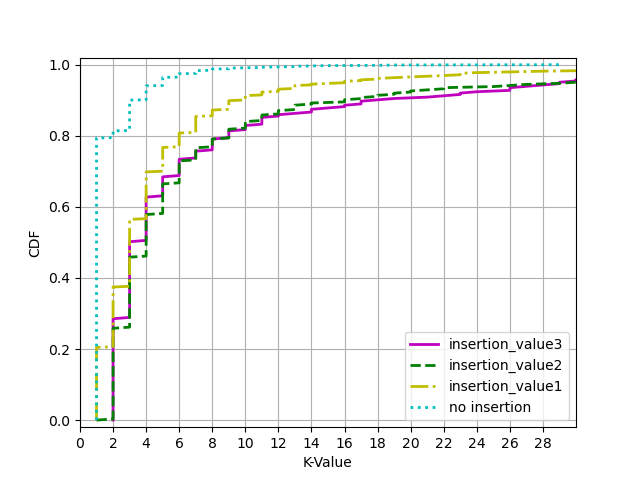}
	\caption{K-anonymity values for different insertion factors and a sharding value of 2.}
	\label{fig:insertion_sharding1}
\end{figure}



Figure~\ref{fig:insertion_sharding1} shows the CDF of k-anonymity values for different insertion factors (no insertion and insertion factors of 1, 2 and 3) and sharding value of 2. The figure shows that, even for the lowest sharding value, an insertion factor of 2 results in k-anonymity larger than 2 for over 98\% of users. Insertion factors of 1, 2 and 3, yield k-anonymity larger than two for 79\%, 98\% and 100\% of users respectively. With a sharding value of 8 (not shown), increasing the insertion values by the same degree yields 2-anonymity for 86\%, 99\% and 100\% with insertion factors of 1, 2 and 3.

\begin{figure}[ht]
	\centering
	\includegraphics[width=0.8\columnwidth]{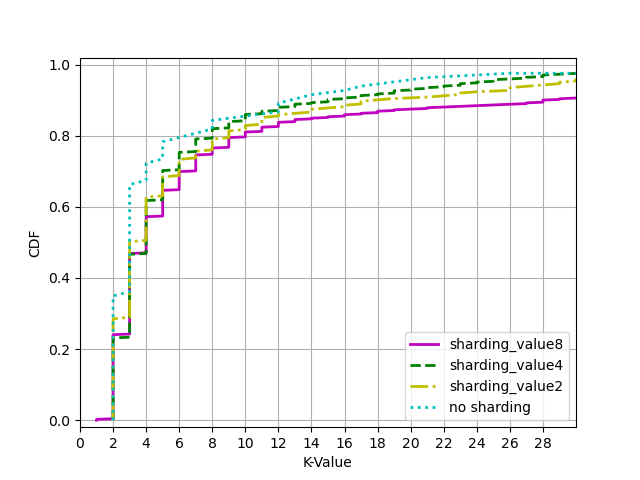}
	\caption{K-anonymity values of users in our dataset with insertion factor of 3.}
	\label{fig:insertion_sharding2}
\end{figure}

Figure~\ref{fig:insertion_sharding2} shows the benefit of higher sharding values for a given insertion factor. For an insertion value of 3, sharding values of 2 and 8 increase the number of users with 2-anonymity to 100\% (and 4-anonymity by 53\% of users).

The insertion of popular requests comes at the cost of additional requests. Even for a fixed insertion factor, the number of additional requests inserted will vary as this is a function of the \textit{uniqueness} of a user's request stream. We estimated the mean overhead (and standard deviations) of different insertion factors for users in our dataset. Table~\ref{tab:privacy_overhead} summarizes this overhead,
as percentage of bytes added to a user's request stream, for three different insertion factors -- 1, 2 and 3. For these factors, the average overhead ranges from 20.27\% to 42.29\%. Given that DNS is responsible for less than 0.1\% of total traffic volume~\cite{schomp:harmful}, even the highest overhead should minimally impact network load.

\begin{table}[h!]
	\footnotesize
		\caption{Privacy overhead for insertion factors. }	
	\centering
	\begin{tabular}{| l | c | c |} 
		\hline
		\textbf{Insertion Factor} & Mean (\%) & Std Dev \\
		\hline
		1 & 20.27 &  3.91\\
		2 & 33.37 &  5.33\\
		3 & 42.29 &  5.89\\
		\hline
	\end{tabular}
	\label{tab:privacy_overhead}
\end{table} 

\textit{\name{} uses a combination of popular request insertion and domain specific sharding} to achieve k-anonymity for a large number of users, with relatively low overhead, while avoiding other potential costs of centralization. Figure ~\ref{fig:insertion_sharding2} shows the combined effect of sharding and insertion in the distribution of k-anonymity values across all users in our dataset for an insertion factor of 3. Sharding over 4 resolvers with an insertion factor of 3 yields 2-anonymity for over 100\% of users (and 4-anonymity for $\approx$60\% of users).


\subsection{Performance and Third-party DNS}
\label{subsec:performance}

Despite the global scale of many of these DNS services' infrastructures~\cite{calder:mapping}, not all regions are equally covered by all services. For specific locales, a particular third-party DNS service may not have nearby servers to offer users, or have fewer or farther away servers than a different service. This can negatively impact both DNS resolution times and web QoE~\cite{otto2012content} for users of the many CDNs that continue to rely on DNS for replica selection, translating into lower engagement~\cite{kirshnan:quasi} and potential loss of revenue~\cite{whitenton:needspeed}.

To illustrate the extent to which the \textit{best} DNS service is location specific, we measure the DNS resolution times with different DNS services in different countries.  We select Argentina, Germany, India and the US; one country in each of the top four continents in terms of Internet users (Asia, North America, Europe, South America). In each locale, we determine the performance of three popular services -- Google, Quad9 and Cloudflare -- as well as two regional DNS services. For each country, we select the top50 sites and issue multiple DNS requests (three) for each of the resources included in those pages. We record the minimum TTFB from the three repeated runs to ignore any transient spikes in the latency. To ensure we are assigned to the correct resolver for clients in each specific country, we use a VPN (NordVPN) with servers in the selected countries. 

\begin{figure}[ht]
	\centering
	\subfigure[IN\label{fig:INResolution}]{\includegraphics[width=0.49\columnwidth]{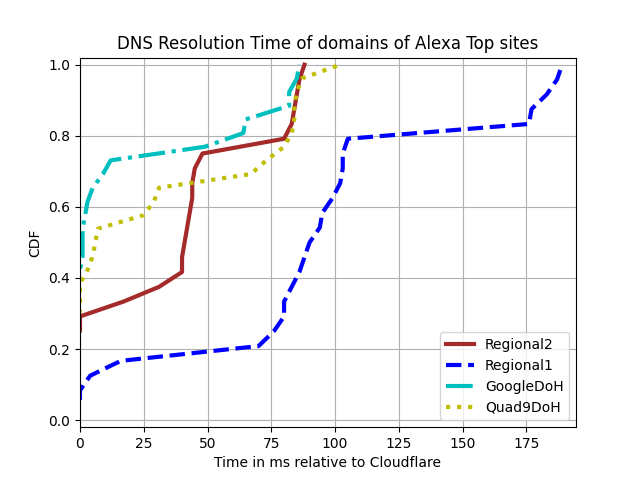}}	
	\subfigure[US\label{fig:USResolution}]{\includegraphics[width=0.49\columnwidth]{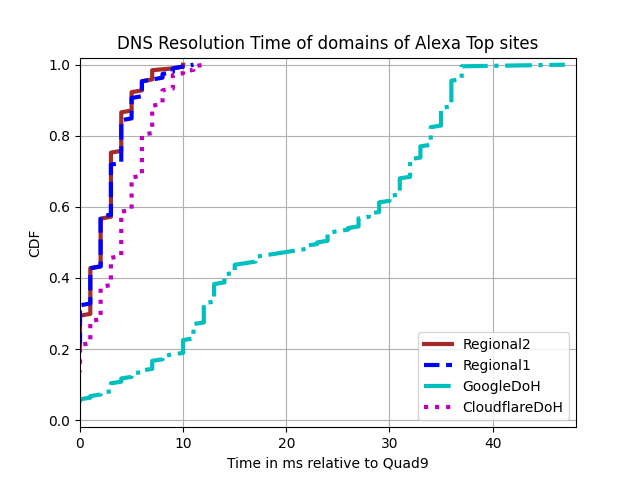}} 		
	\caption{DNS resolution times of individual, public DoH and DNS providers in IN and the US relative to the best performing DNS service in each country (Cloudflare for IN and Quad9 for the US).}
	\label{fig:resolutiontime}
\end{figure}

Figure~\ref{fig:resolutiontime} plots the CDFs of the resolution time for popular resources in two of these countries, India and the US (full set of graphs in \S\ref{sec:appendix}). The resolution times of  all resources, for each DNS provider in a country, are plotted relative to the DNS service performing best in the average case. Service resolution times are relative to Quad9 in the US and to Cloudflare in India. We can see that in the median case, Google performs as well as Cloudflare but Regional2 performs ~40ms worse than Cloudflare and Regional1 performs 87ms worse than Cloudflare in India. In the US, Regional 1, Regional2 and Cloudflare have a difference of less than 10ms from the best resolver(Quad9) whereas Google performs ~25ms worse than Quad9(for the 50th percentile).

\begin{figure}[htb!]
	\includegraphics[width=0.9\columnwidth]{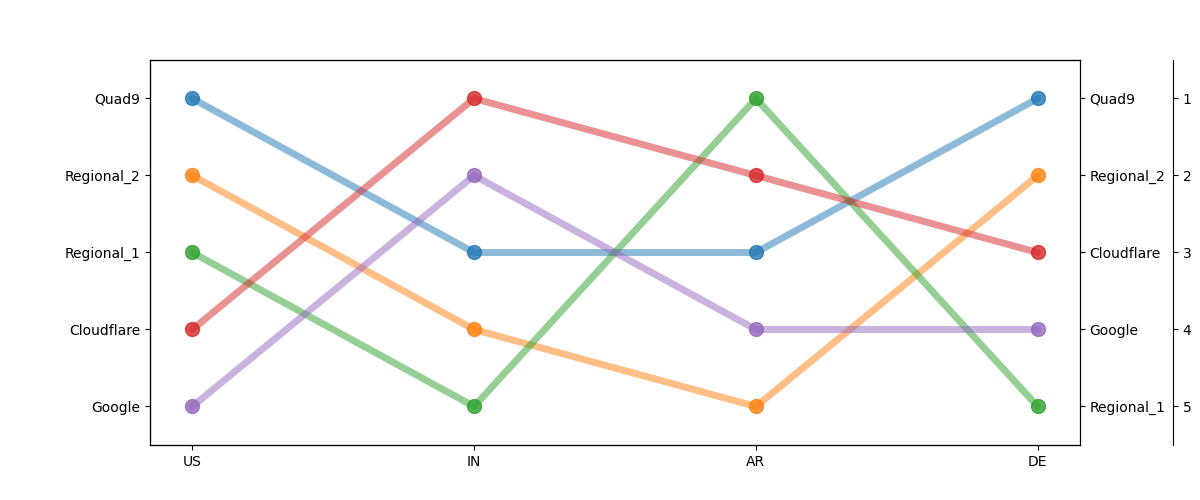}
	\caption{A bump graph of DNS resolution times ranking across countries (Regional DNS services in Table~\ref{tab:regional_ip}). The maximum (avg) Kendall tau distance between rankings is 0.6 (0.37) showing that there is no \textit{best} DNS service for every location.}
	\label{fig:resolutiontimeranking}
\end{figure}

Figure~\ref{fig:resolutiontimeranking} is a bump graph that summarizes the rankings of the evaluated services, based on DNS resolution times, across countries. Each service is associated with a differently colored line. The different lines change positions as the associated service changes rank positions between countries. The plot clearly illustrates that the ranking among DNS services varies widely across locales\footnote{We repeated this measurement over a period of a week and noticed that the ranking changed over time as well} . Figure~\ref{fig:resolutiontime} illustrates the differences in relative resolution times using Germany and the US as examples (full set of graphs in \S\ref{sec:appendix}). While Quad9 has the best resolution times in the US, Google ranks the lowest. On the other hand, Cloudflare has the best resolution times in India while Google performs second to the best. The maximum (average) Kendall tau distance\footnote{The Kendall tau rank distance is a metric, ranging form 0 to 1, that counts the number of pairwise disagreements between two ranking lists.} between rankings is 0.6 (0.37) clearly showing that there is no \textit{best} DNS service for every location.

\begin{table}[h!]
	\footnotesize
	\centering
		\caption{Regional DNS resolvers IP references. }
	\begin{tabular}{| l | c | c |} 
		\hline
		\textbf{Countries} & Regional 1 & Regional 2 \\
		\hline
		AR & 190.151.144.21 & 200.110.130.194 \\
		DE & 46.182.19.48 & 159.69.114.157 \\
		IN & 182.71.213.139 & 111.93.163.56 \\
		US & 208.67.222.222 & 208.67.220.220 \\
		\hline
	\end{tabular}
	\label{tab:regional_ip}
\end{table}

\begin{figure}[ht]
	\centering
	\subfigure[DE\label{fig:DEResolution_sharding}]{\includegraphics[width=0.8\columnwidth]{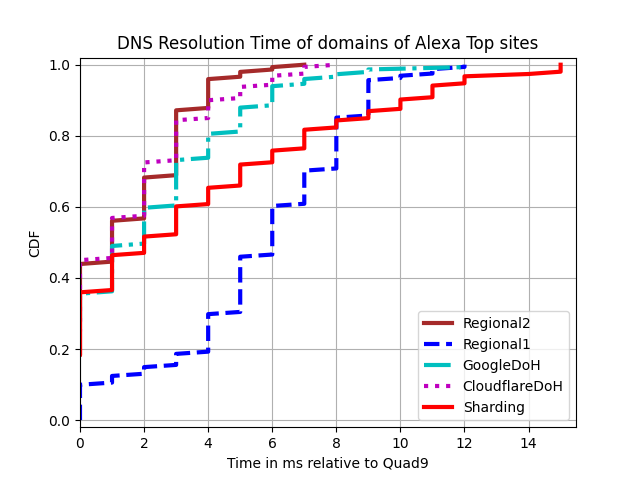}}
	\caption{DNS resolution times of individual, public DoH and DNS providers and sharding in Germany relative to the best performing DNS service (Quad9).}
	\label{fig:resolutiontime_sharding}
\end{figure}

The differences in performance across DNS services means that sharding requests will result in high performance variability. To illustrate this we plot CDFs of the resolution time for popular resources in different countries using different DNS services and basic sharding.  Figure~\ref{fig:resolutiontime_sharding} shows the resolution times for each service relative to the best performing resolver in Germany (full set of graphs in \S\ref{sec:appendix}). The figures clearly show that the resolution times of sharding in each locale are affected by the worst performing DNS service in a locale (Regional1 service in Germany) thereby increasing the resolution times of sharding in comparison to the better performing DNS services.

Domain specific sharding, while improving privacy, will most likely not improve performance variability as all queries belonging to a given domain could be assigned to a randomly selected (potentially the slowest/fastest for the locale) DNS service and the association persists for all resolution request of the associated domain. 

\subsubsection{Racing}
\label{subsubsection:racing} 

\name{} addresses this potential problem by using resolution races~\cite{vulimiri:racing} that place DNS services in competition with each other. For each DNS request, \name{}~uses a configurable number of DNS resolvers and issues the same request to each of them simultaneously, returning to the client the first response from the first resolver that successfully resolves the query addressing the problem of high latency caused by latency variance. This redundancy comes with negligible cost on bandwidth for potentially significant improvements on resolution times. 

A potential issue of racing DNS services is the associated tradeoff between privacy and performance. Racing a larger number of DNS services increases the chances of finding the \textit{best} resolver for that request and locale, but also increases the fraction of the user's DNS request stream seen by any particular service. 

To illustrate this tradeoff, we conduct an experiment measuring the performance and request stream exposure of different racing configurations. We shard 100 resources, in the top50 sites in the US, across different DNS providers. We issue resolution requests for these resources to a randomly selected resolver, and run races between two and three resolvers for each resolution request. Table~\ref{table:fraction_of_requests} presents a summary of these results.  


\begin{table}[htb!]
	\centering
		\caption{Resolution times with \name{} running no resolution races, racing 2 and racing 3 resolvers, and the associated percentage of requests each resolver receives. 
	}
	\begin{tabular}{|c|c|c|c|c|}
		\hline
		& \multicolumn{2}{|c|}{Resol Time} & \multicolumn{2}{|c|}{Request \%}\\
		\hline
		& \textbf{Mean} & \textbf{Std Dev} &\textbf{Max} &\textbf{Avg} \\
		\hline
		No Racing & 42.22 & 38.13 &20.34&12.5\\
		Racing 2& 25.83 & 12.42&28.10&25.0 \\
		Racing 3& 21.34 & 6.53&42.31&37.5 \\
		\hline
	\end{tabular}
	\label{table:fraction_of_requests}
\end{table}

The table show the performance advantages of racing. Just racing two resolvers offers a percentage improvement of 38.8\%, while racing three provides an additional 17.88\% improvement on resolution time.\footnote{A reminder of \textit{the power of two random choices}~\cite{mitzenmacher:2choice}.} 

Table~\ref{table:fraction_of_requests} also show the tradeoff posed by increasing the number of raced resolvers showing the percentage increase (maximum and average) of the DNS request stream exposed by each of these strategies to the different resolvers. The maximum percentage of requests seen by each resolver increases from 20.34\%, with \textit{no racing}, to 28.10\% when racing 2 resolvers and 42.31\% when racing 3 resolvers. Racing 2 resolvers achieves the best balance between privacy and performance in our analysis and we set this as the default configuration of \name.

\begin{figure}[h!]
	\centering
	\includegraphics[width=0.85\linewidth]{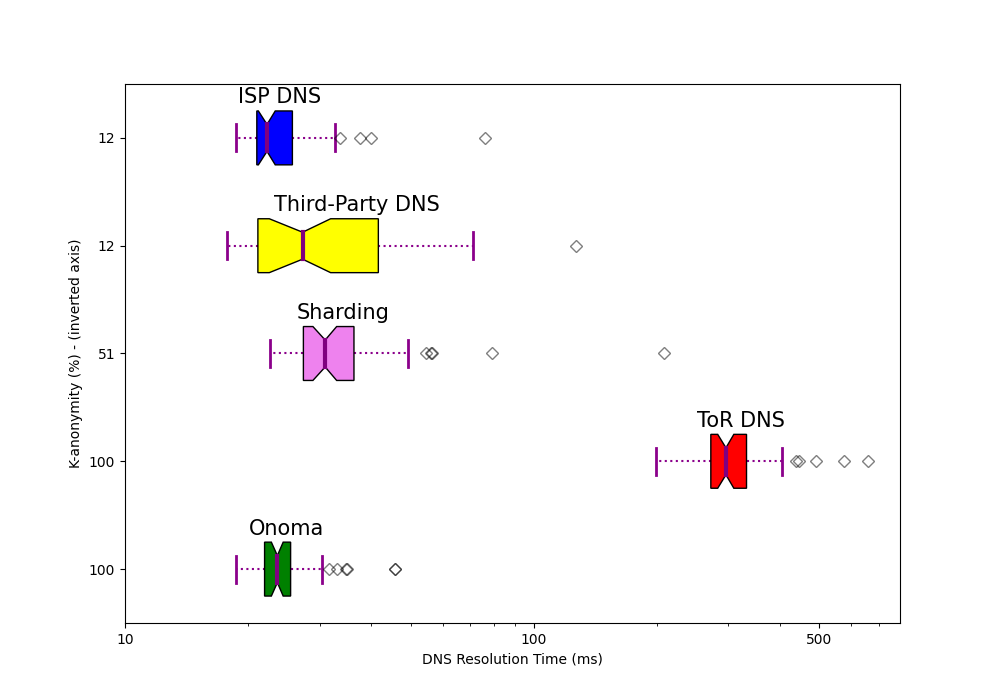}
	\caption{Privacy and/or Performance. Average DNS resolution times and corresponding 2-anonymity values of different DNS variants and \name.}
	\label{fig:perfvsprivacy}
\end{figure}

Finally, we also show the tradeoff between privacy and performance across different DNS variants, such as DNS over Tor, our ISP DNS, third-party DNS services, sharding and \name. Figure~\ref{fig:perfvsprivacy} shows box plots of resolution times (x axis) for each DNS variant and the y-axis shows the privacy gain for each DNS variant in terms of the percentage of users that have at least 2-anonymity. The figure shows that DNS over Tor gives the maximum privacy but significantly worse resolution times, whereas ISP DNS while providing the best performance compromises on privacy significantly. We show that sharding improves 2-anonymity but by randomly distributing requests across popular DNS providers it can also significantly hurt performance. \name, on the other hand, gives the best privacy and performance for users.

\subsubsection{Direct Resolution}
\label{subsubsection:directresolution} 

Even if certain third-party DNS services result in faster resolution time for their users, the same users can still experience worsening Web QoE, depending on the specific deployment of the DNS services and popular CDNs infrastructures in their region, and the mapping policies of these CDNs. Differences in relative distance between the users and their recursive DNS resolvers can result in poor mapping of a user to a CDN replica server and higher delays in getting objects hosted by those replicas. 

\begin{figure}[h!]
    \centering
    \includegraphics[width=0.9\linewidth]{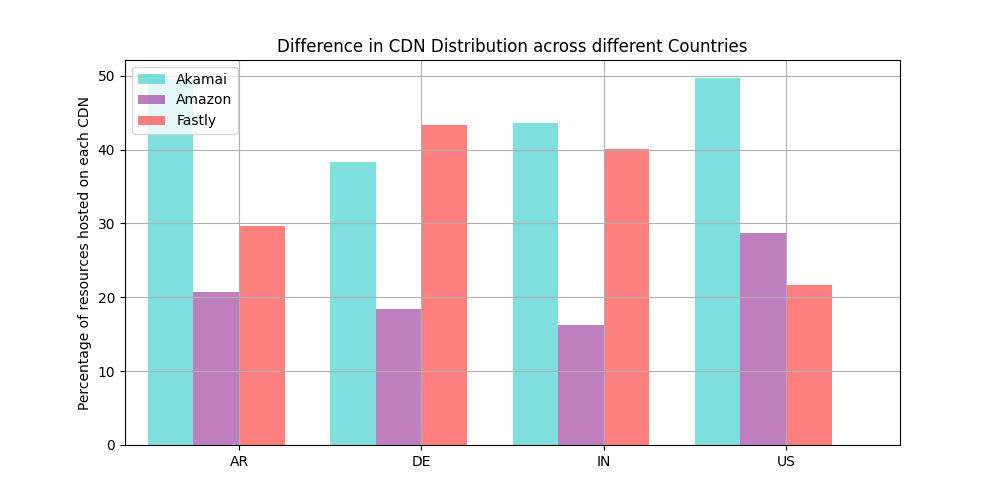}
    \caption{Relative popularity of CDNs in top50 Regional sites in Argentina, Germany, India and the US.}
    \label{fig:CDNDistribution}
\end{figure}



\textbf{The popularity of CDNs varies across locales.} Figure~\ref{fig:CDNDistribution} shows
the relative popularity of different CDNs per country. Each cluster represents a country and each bar indicates the number of resources, from the top50 regional sites, hosted on a particular CDN. Clearly, the relative ranking of CDNs as host to the most poular resources varies across locales. For instance Germany's popular sites and resources seem to be preferentially hosted on Fastly (44\% resources) whereas, the other three countries rely primarily -- although to different degrees -- on Akamai (ranging from 44\% to 50\%). Additionally Amazon has relatively better presence in the US, but not in the other three countries. 

\begin{figure*}[ht]
    \centering
    \subfigure[AR,Akamai\label{fig:AR_akamai_ttfb}]{\includegraphics[width=0.24\textwidth]{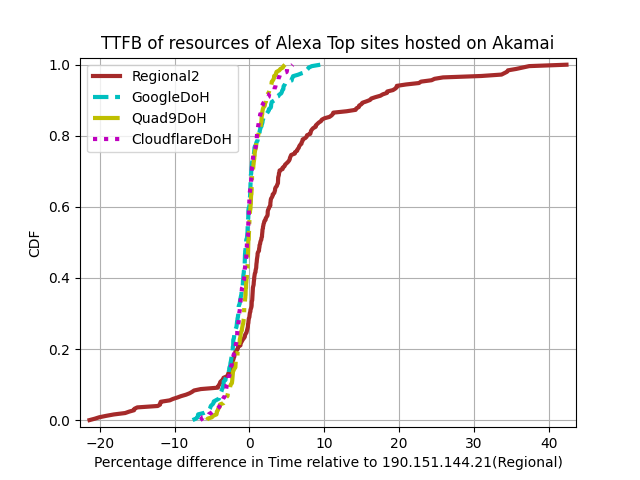}}
    \subfigure[DE,Fastly\label{fig:DE_fastly_ttfb}]{\includegraphics[width=0.24\textwidth]{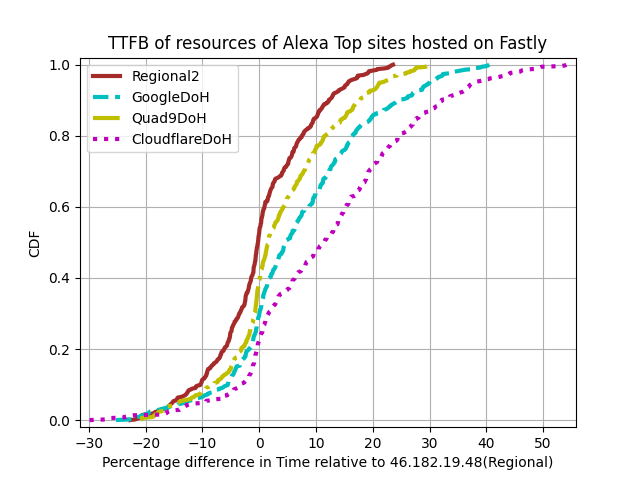}}
    \subfigure[IN,Akamai\label{fig:IN_akamai_ttfb}]{\includegraphics[width=0.24\textwidth]{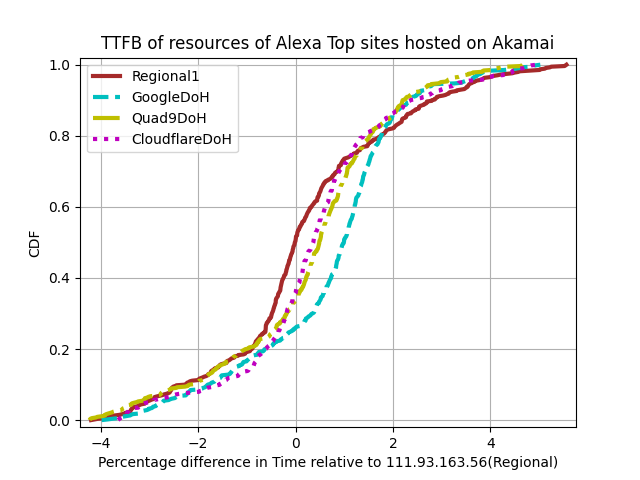}}
    \subfigure[US,Akamai\label{fig:US_akamai_ttfb}]{\includegraphics[width=0.24\textwidth]{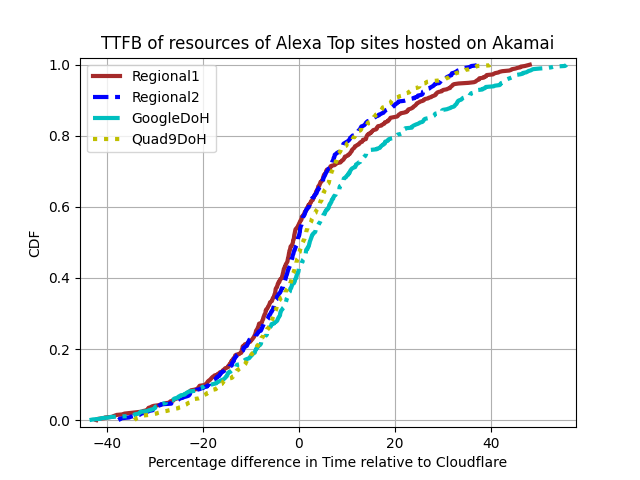}} 
    \newline
    \caption{Percentage Difference in TTFB of individual, public DoH and DNS resolvers per country, relative to the best DNS service in that locale, across different countries. Cloudflare which performs worst in Germany for content hosted on Fastly, shows 10\% greater TTFB in the median case compared to the best service (Regional).}
    \label{fig:ttfb}
\end{figure*}

\begin{figure}[ht]
	\centering
	\subfigure[Akamai\label{fig:ttfbRankingAkamai}]{\includegraphics[width=0.8\columnwidth]{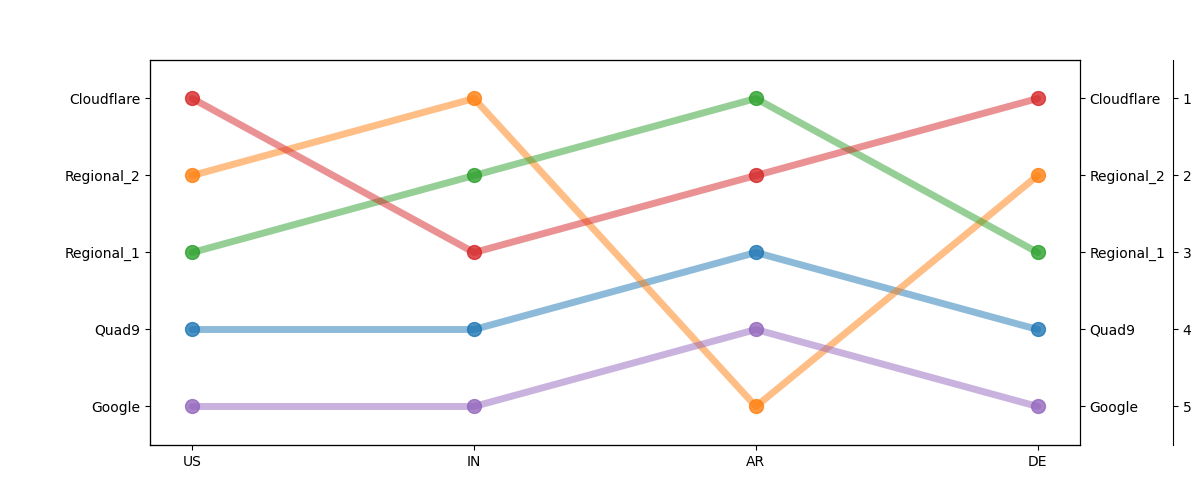}} 
	\subfigure[Amazon\label{fig:ttfbRankingAmazon}]{\includegraphics[width=0.8\columnwidth]{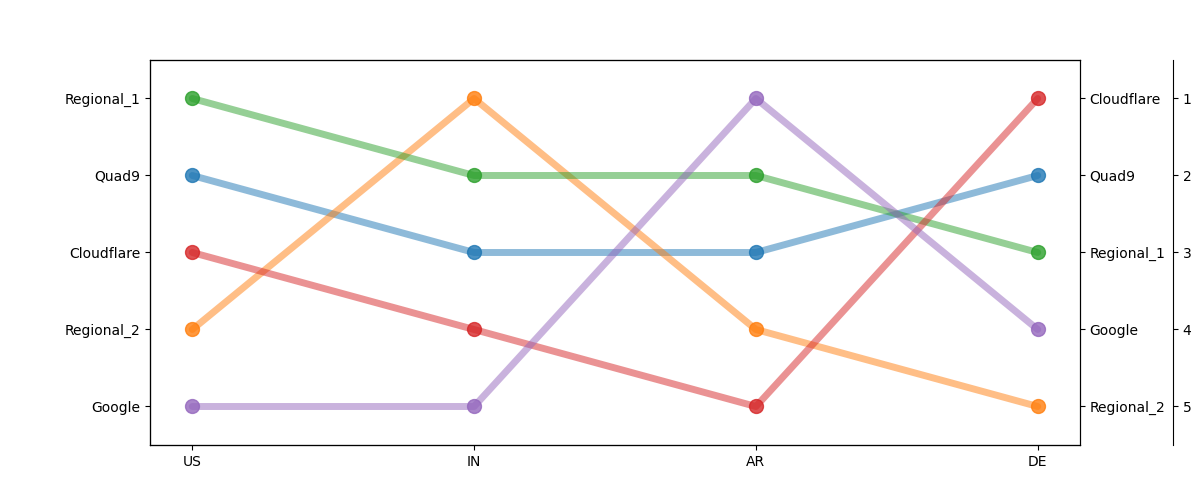}}
	\caption{DNS resolvers ranking by TTFB for resources hosted on different CDN across different countries.
		The maximum (avg) Kendall tau distance between rankings is 0.4 (0.27) for Akamai and 0.7 (0.53) for Amazon.}
	\label{fig:ttfbRanking}
\end{figure}

\textbf{User's QoE is impacted by the choice of DNS service depending on the locale.} To evaluate how users' QoE in different locales for content hosted on different CDNs varies with the choice of DNS service, we then conduct an experiment measuring time-to-first-byte (TTFB) as a proxy for Web QoE. TTFB gives the minimum response time for the content obtained from the CDN replica and thus measures the effectiveness of the CDN mapping. For this experiment, we use resources from the top50 regional sites hosted on the most popular CDN in each country (Fig.~\ref{fig:CDNDistribution}).   

We load each resource three times using Google's lighthouse~\cite{lighthouse} for different (CDN, country) combinations using the same VPN service and record the minimum TTFB for each resource. We use the minimum value to ignore any transient spikes in TTFB measurements. 

From the recorded TTFB measurements with each resolver for the (CDN, country) combination, we find the best performing resolver, in the median case, and calculate the percentage difference in TTFB of each DNS service from this baseline for the given CDN and locale.

Figure~\ref{fig:ttfb} plots a CDF of this difference in TTFB. The figure shows that different DNS providers result in different TTFB for resources hosted on different CDNs, across different locales. Cloudflare which performs worst in Germany for content hosted on Fastly, shows 10\% higher TTFB in the median case compared to the best service (Regional1) and Google (with the worst TTFB in IN,Akamai) shows 1\% higher TTFB in the median case compared to the best service (Regional2). 

Figure~\ref{fig:ttfbRanking} is a bump graph further highlighting this point (full set of graphs in \S\ref{sec:appendix}). The figure ranks DNS resolvers by TTFB for resources hosted on different CDNs across different countries and highlight their relative changes in ranks. For instance, Google gives the worst QoE for content hosted on Akamai in the US and India (even though Google gives second best DNS resolution times in India as shown in Fig.~\ref{fig:resolutiontimeranking}), but results in best QoE for content hosted on Amazon in Argentina (where it has the second worst resolution time as seen from Fig.~\ref{fig:resolutiontimeranking}). This reiterates that the user's web QoE in fact depends on the specific deployment of the DNS services and popular CDNs infrastructures in their region, and the mapping policies of these CDNs.

\textit{To address the QoE impact of using different third-party DNS resolvers and CDNs in different locales, \name~combines racing with direct resolution~\cite{otto2012content} when resolving content that is hosted by a CDN.} It uses the response from the fastest DNS resolver for direct resolution. Direct resolution combines the best aspect of interative and recursive DNS resolution to obtain improved CDN redirections for each resource.  It leverages the cache of a recursive
resolver to efficiently map a Canonical Name (CNAME) to an authoritative name server, but directly contacts the CDN's
authoritative server to obtain a precise redirection for the client. This gives the CDN's authoritative server better information of the clients' actual network location and allows for a better replica assignment~\cite{otto2012content}. 

To understand the aggregated benefits of the combined approach, we compare the performance of \name{} with that of sharding. Since sharding, by itself, randomly distributes DNS requests across a set of public DNS servers, the resulting performance matches the average performance of public DNS services in each region and can serve as a baseline for comparison. We load each resource, of the regional top50 sites three times and compute the minimum TTFB. 

\begin{figure}[htb!]
	\centering
	\subfigure[AR,Akamai\label{fig:AR_OnomavsProxyAkamai}]{\includegraphics[width=0.4\textwidth]{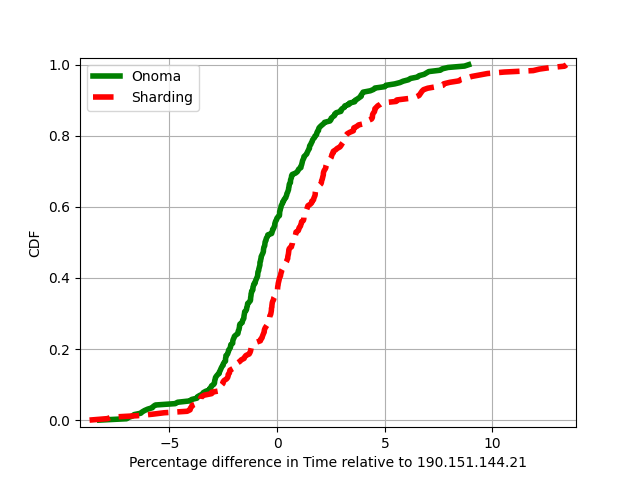}}
	\caption{Percentage Difference in TTFB of Sharding and \name{} relative to the best DNS service across the US and Argentina.}
	\label{fig:OnomavsProxy1}
\end{figure}

Figure~\ref{fig:OnomavsProxy1} presents a CDN of the percentage difference in TTFB for both \name~and sharding, relative to the best performing DNS service for a particular (CDN, country) combination (complete graphs in \S\ref{sec:appendix}). The figure includes plots for Argentina for content hosted on Akamai. The graphs show the clear dominance of \name~over sharding across CDNs in both locales, with a median improvement of 9.34\% in Argentina with Akamai CDN.


\subsection{Overview}

\name~lets users leverage the benefits of third-party DNS services while $(i)$ providing improved k-anonymity, via a combination of sharding and popular domain insertion, $(ii)$ consistently better resolution times than individual DNS resolvers, independently of the locale, by running resolution races, and $(iii)$ and without impacting users' web experience via direct resolution. 

\begin{figure}[h]
	\centering
	\includegraphics[width=0.7\linewidth]{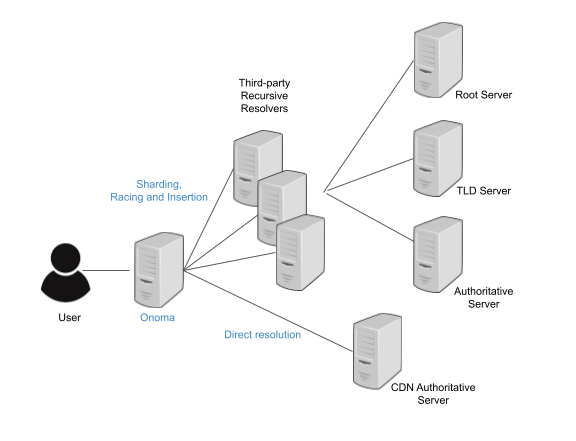}
	\caption{\name~interaction with DNS resolvers.}
	\label{fig:model}
\end{figure}

Pseudocode~\ref{algo:code} provides a high-level description of \name's DNS request resolution process, while Fig.~\ref{fig:model} illustrates its interaction with the different DNS resolvers.

\begin{algorithm}[htb!]
\caption{Resolution process followed by \name.}	\label{algo:code}
\begin{algorithmic}\scriptsize
	\For{DNS resolution}
	\If{domain $\notin$ DomainMap} 
	\State Run races among resolution 
	\State Add domain and winner resolver to DomainMap
	\EndIf
	\If{domain $\notin$ popular domains}
	\For{insertion-factor}
	\State Sample a popular domain 
	\State Insert the domain in request stream
	\EndFor
	\EndIf
	\If{CDN-ized domain}
	\State Direct resolution for domain
	\EndIf 
	\EndFor
	
\end{algorithmic}
\end{algorithm}

\section{Implementation}
\label{sec:implementation}

We have implemented \name{} as an end-system resolver running on users' machines. \name{} is implemented in 3,510 lines of Go code, excluding comments, and is designed as a background service. It is readily deployable and easy-to-install on any operating system and neither requires any modification to the user's operating system nor coordination with existing CDN or DNS services. The implementation includes a light-weight, client-side application (under 500MB) hosting a simple user interface to let users install, start and stop the service and modify default configuration settings such as changing the set of DNS resolvers used for sharding and racing, and the request insertion factor. \name{} maintains a set of DNS services which includes popular public encrypted services -- Google, Cloudflare and Quad9 -- as well as a set of regional public DNS services available at the clients' location. 

\name{} also implements a measurement module to perform experimentation on the clients' side. \name{} uses the collected results to indentify optimal default values for sharding, racing and insertion, as well as dynamically select the best DNS resolvers for the user's locale. 



\section{End-to-end Evaluation}
\label{sec:evaluation}

In this section, we present results from an end-to-end evaluation of \name, combining all the design features discussed. In our evaluation we compare \name's performance with that of different third-party DNS services, across locales and CDNs. We use TTFB, as the performance metric for comparison, and the resources hosted on the top50 regional sites, as a proxy of users' QoE in those regions.

The reported results were collected by the measurement module that is part of \name{}, which runs all the experiments in the background (\S\ref{sec:implementation}). We set up the evaluation measurements in the selected locations using a VPN service, using different DNS settings including public DoH resolvers, regional resolvers and \name{}. We focus on three CDN services -- Akamai, Amazon and Fastly -- and the resources they host.

For each DNS setting and (CDN, country) combination, the measurement module loads the set of resources with Google Lighthouse. It records the minimum TTFB from three repeated runs, finds the best performing resolver, in the median case, and calculates the percentage difference in TTFB between each DNS service and the indentified best DNS service for a given CDN and locale. 

\begin{figure*}[ht]
    \centering
    \subfigure[AR,Akamai\label{fig:AR_akamai_ttfb}]{\includegraphics[width=0.27\textwidth]{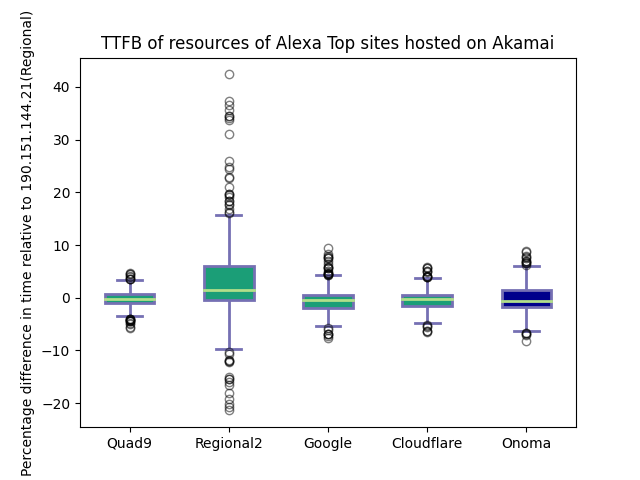}}
    \subfigure[IN,Akamai\label{fig:IN_akamai_ttfb}]{\includegraphics[width=0.27\textwidth]{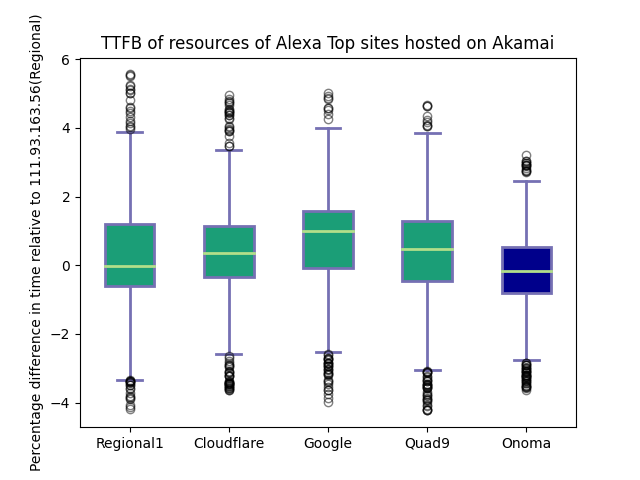}}
    \subfigure[US,Akamai\label{fig:US_akamai_ttfb}]{\includegraphics[width=0.27\textwidth]{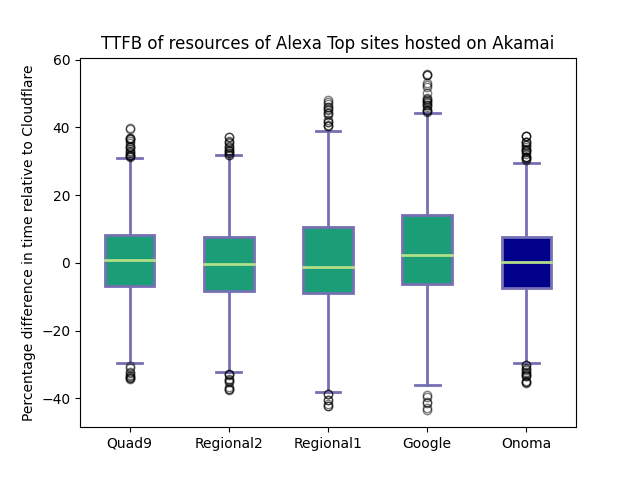}}
    \newline
    \subfigure[DE,Akamai\label{fig:DE_akamai_ttfb}]{\includegraphics[width=0.27\textwidth]{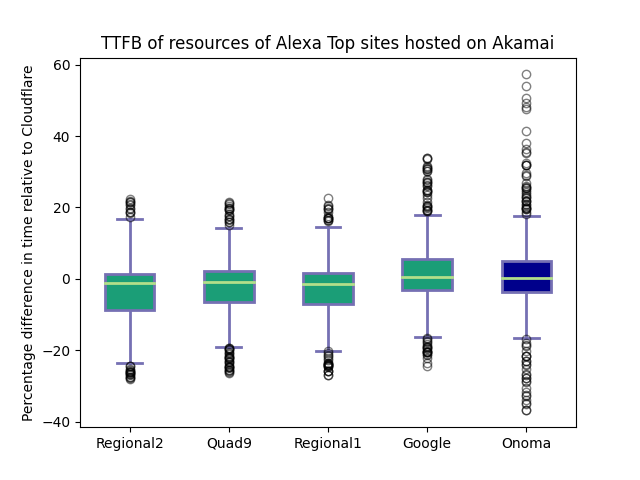}}
    \subfigure[DE,Amazon\label{fig:DE_amazon_ttfb}]{\includegraphics[width=0.27\textwidth]{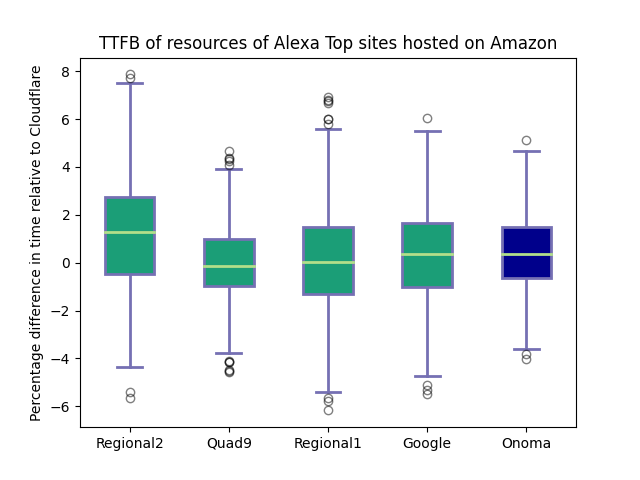}}
    \subfigure[DE,Fastly\label{fig:DE_fastly_ttfb}]{\includegraphics[width=0.27\textwidth]{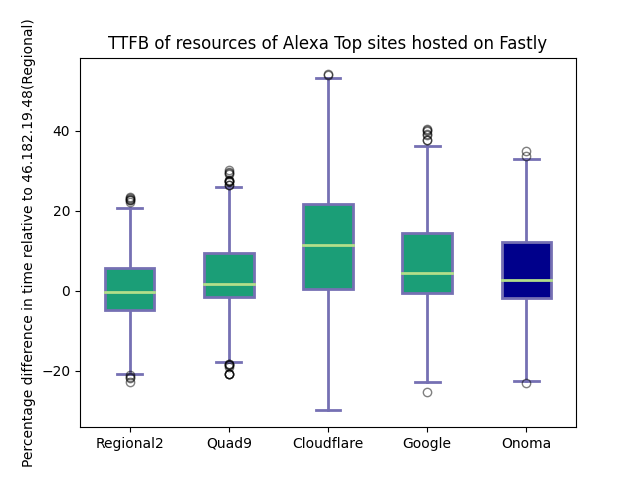}} 
    \newline
    \caption{Performance of \name~and of individual, public DoH and DNS resolvers per country, relative to the best DNS service across CDNs and countries. The top set focuses on one CDN, Akamai, across countries while the bottom set focuses on different CDNs in Germany. The graph for Akamai in Germany is included only once. \name~offers consistently better performance across CDN and country combinations.}
    \label{fig:ttfb_wOnoma1}
\end{figure*}

Figure~\ref{fig:ttfb_wOnoma1} presents boxplots of the performance of different DNS services and \name{} (full set of graphs in \S\ref{sec:appendix}). The plots are relative to the best performing resolver for a given CDN and locale (e.g., Cloudflare for Akamai in the US). Boxplots clearly show the median and the first and third quartiles, as well as the minimum and maximum values (upper and lower whiskers) of the collected TTFB data. The outliers of the data are shown by circles above and below the whiskers.

The figure shows that the performance of \name{} is either best, or comparable to the average performance of the public DNS services. \name{} gives consistently a lower inter-quartile range (third quartile - first quartile), indicated by the colored boxes in the plot (blue box for other providers and green for \name{}), for most cases. Median performance of \name{} is better than the best Service for most CDN, country combinations or at least comparable to all the DNS services tested for that locale. We see significant performance improvement with content hosted on Amazon in Germany and with content hosted on Akamai in India and the US. For instance, the inter-quartile range (IQR) of \name{} is 18.4\% in the US, for Akamai's content, but for other DNS services it ranges between 18.7\% to 24.8\%. For Akamai's content in India, \name{} gives an IQR of 1.7\% and for other DNS services the IQR ranges between 1.7\% to 2.5\%. Note that \name{} attains this for any combination of CDN and locale and with high values of k-anonymity (\S\ref{subsection:privacy}). 




\section{Related Work}
\label{sec:related_work}

Our work builds on an extensive body of research on characterizing and improving DNS performance, privacy and its impact on users' QoE. For over two decades since Mockapetris~\cite{mockapetris:rfc1034}, DNS has been the subject of several measurement studies in the wild~\cite{ager:dns, huitema:dnsperf, wills:dnsweb, jung:dnscaching, liston:dnsperf, mao:proximity, shaikh:selection, otto2012content}. Much of this work has focused on caching, performance and scalability across locales (e.g., \cite{jung:dnscaching, liston:dnsperf}, and the interaction between DNS and content distribution (e.g., \cite{mao:proximity,shaikh:selection,otto2012content}). 

The appearance of open DNS resolvers, offered by companies such as Google and OpenDNS, and the follow-up EDNS-client-subnet DNS extension (ECS)~\cite{contavalli:rfc7871}, motivated prior work to understand these services' relative performance~\cite{ager:dns} and their potential impact on QoE~\cite{otto2012content,pujol:steering,fan:ecs}. These and other related efforts have shown that user-mapping accuracy is worsening, leading to increased delays, lower engagement~\cite{kirshnan:quasi} and revenue loss~\cite{whitenton:needspeed}. 

Despite some adoption of anycast~\cite{calder:anycastcdn,alzoubi:ayncastcdn} and the promise of CDN-ISP collaborations~\cite{pujol:steering,poese:padis,xie:p4p,alimi:alto}, many CDNs continue to rely on DNS-based server redirection. Otto et al.~\cite{otto2012content} proposed \textit{direct resolution} as an alternative end-user approach to address the impact of remote DNS resolvers on Web QoE despite the limited adoption of ECS. Direct resolution leverages the recursive DNS server to obtain the authoritative server for CDN-hosted content, but have the client itself do the final resolution by directly contacting the authoritative server operated by the CDN. This yields better localization and, thus QoE performance, without additional privacy costs~\cite{otto2012namehelp}. \name~adopts direct resolution, including caching of the CDN-run authoritative name server to avoid additional delay in resolution (Sec.\ref{sec:system_design}).

Beyond performance, increasing concerns about user privacy has served as motivation for new DNS techniques and services. Recent work evaluates the relative performance and QoE impact of DNS-over-Encryption variants~\cite{bottger:dohqoe,hounsel:enccost}, and assess their implementations, deployment configurations~\cite{liu:doe,deccio:dnsprivacy} and policy implications~\cite{dns:dprive,edns:policy}. Deccio et al.\cite{deccio:dnsprivacy} also show that only 0.15\% of open resolvers and a negligible number of authoritative servers support DNS privacy, and those that do comes from a handful of popular DNS providers such as Cloudflare, Facebook, Google, Quad9. 

Although the benefits provided by DNS-over-Encryption against certain threats are clear, there are significant privacy concerns associated with the need to trust the DNS operator with the entire request trace of users. Schmitt et al.~\cite{schmitt:odns} proposes \textit{Oblivious DNS} to address centralization, by obfuscating the queries that the recursive resolver sees by adding its own authoritative ODNS resolver. A few closely related efforts~\cite{hoang:sharding,hounsel:ddns,jari:sharding} have proposed and evaluated the use of DNS request sharding to mitigate the privacy concerns from exposing all DNS resolution to a single DNS-over-Encryption service, potentially grouping requests per domain to avoid distribution of similar names to different resolvers~\cite{jari:sharding}. \name{} builds on this idea (Sec.~\ref{sec:goals}) and shows the degree of additional privacy it provides by avoiding DNS-based user-reidentification through domain specific sharding and by adaptively inserting popular requests on a user DNS request stream based on the uniqueness of a requested domain.  \name{} also avoids the potential performance costs of sharding by dynamically selecting the different DNS services to use, depending on the user's location, and racing resolvers~\cite{vulimiri:racing}. Other recent works have questioned the actual privacy benefits of encrypted DNS and proposed solutions to address some of them (e.g.,~\cite{siby:encrypted,bushart:padding}) which could be easily adopted by \name{}. 

The design of \name, leverages an extensive body of work to introduce, to the best of our knowledge, the first end-user DNS resolver to take advantage of the privacy and performance benefits of centralized DNS services while avoiding the associated costs.



\section{Limitations and Future Work}
\label{sec:limitation}

In this section we discuss limitations of our approach and its evaluation. For starters, we focus on challenging attempts at user re-identification by third-party DNS resolvers, motivated by the findings of Olejnik et al.~\cite{olejnik:browsinghist} and Bird et al.~\cite{bird:browsinghist}. 
Due to the privileged position of DNS resolvers in the resolution path, the best privacy a user could expect is in terms of k-anonymity. We adopt the user identification approach and follow the evaluation model and settings of prior work~\cite{olejnik:browsinghist,bird:browsinghist}.  We do not claim this to be the only approach to (re)identify users, but we posit that the combination of popular domain insertion and sharding will complicate the task of user identification by most third-party resolvers. The opportunistic use of Oblivious DNS~\cite{odns:cloudflare,schmitt:odns} for resolving truly distinctive or otherwise problematic domains further challenge re-identification while limiting the performance impact of the technique.

In this work we opted for Alexa’s regional ranking as this offers country specific rankings available to the community, and includes in its listing regional domain aliases for instance yahoo.com.jp and yahoo.com. Other popular rankings, such as Tranco compile their regional rankings by computing an intersection global ranking and the domains appearing in the country-specific Chrome User Experience Report list. These rankings, therefore, miss all region specific domain aliases. While we are exploring the use similarweb's~\cite{similarweb} regional ranking for future work, we note that our analysis and evaluation of \name~is largely independent of specific website rankings. 

Finally, we also note that the dataset used for our evaluation, collected from our university network for a two-week period, can be biased both in terms of the population and the time window over which it was collected. In future work, we are exploring ways to expand our evaluation dataset.


\section{Conclusion}

Third-party recursive DNS services offer a number of valuable features, from high resolution performance to 
improved privacy through encrypted and Oblivious DNS. At the same time, these DNS variants are operated by 
a handful of providers such as Google, Amazon and Cloudflare, strengthening a concerning trend toward DNS 
centralization and its implications, ranging from user privacy and QoE, to system resilience and market 
competition. We presented {\it \name}, an end-system DNS resolver that let users take advantage of third-party 
DNS services without sacrificing privacy or performance. Our evaluation shows the benefits of \name~across 
locales, with different DNS services, content providers, and CDNs. 

\clearpage

\SuspendCounters{totalpages} 

\bibliographystyle{ACM-Reference-Format}
\bibliography{reference}
\appendix
\clearpage
\section{Appendix}
\label{sec:appendix}

In this section we include the remaining graphs from our motivation, system design and evaluation of \name.

\subsection{The importance of DNS for CDN replica selection}

\begin{figure}[ht!]
	\centering
	\subfigure[Amazon \label{fig:Amazon_DNS_localization_a}]{\includegraphics[width=0.48\columnwidth]{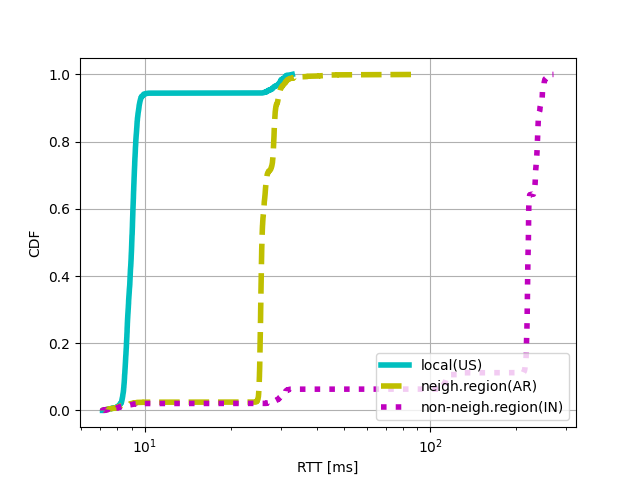}}
	\subfigure[Fastly\label{fig:Fastly_DNS_localization_a}]{\includegraphics[width=0.48\columnwidth]{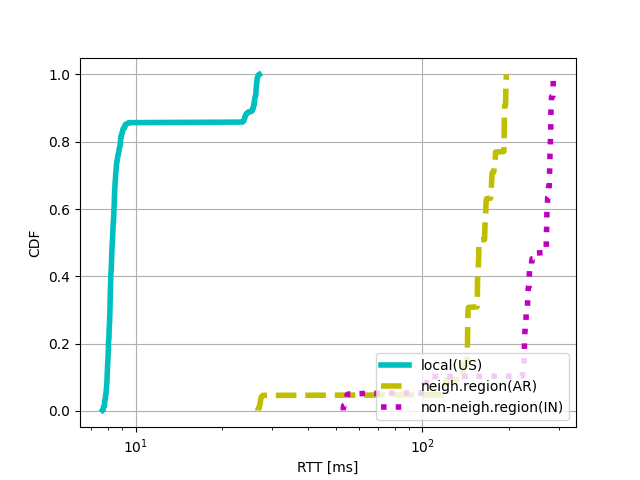}}
	\caption{Latency to assigned replica servers (/24) for Amazon and Fastly for resources of top50 US websites. }
	\label{fig:DNS_localization_appendix}
\end{figure}

Figure~\ref{fig:DNS_localization_appendix} shows the remaining graphs to illustrate the importance of DNS for redirection. The client in the US is using a local DNS resolver (cyan) and distant ones in Argentina and India.

\begin{figure}
	\centering
	\subfigure[Akamai\label{fig:Akamai_DNS_localization_DE}]{\includegraphics[width=0.2\textwidth]{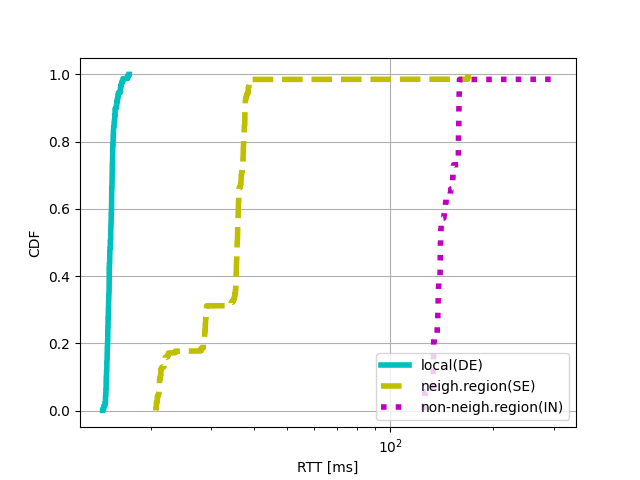}}
	\subfigure[Amazon\label{fig:Amazon_DNS_localization_DE}]{\includegraphics[width=0.2\textwidth]{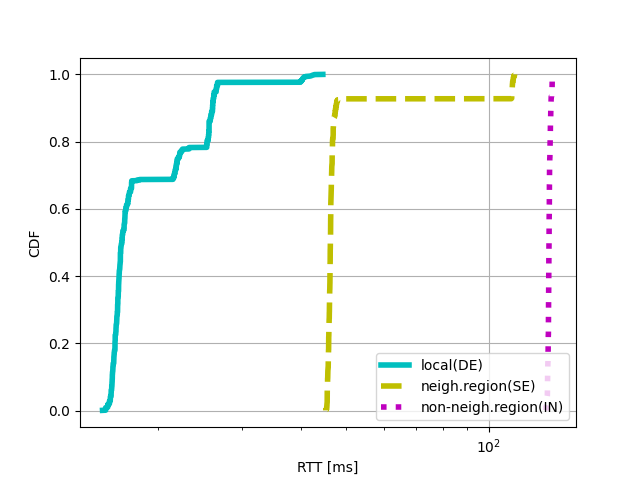}}
	\subfigure[Fastly\label{fig:Fastly_DNS_localization_DE}]{\includegraphics[width=0.2\textwidth]{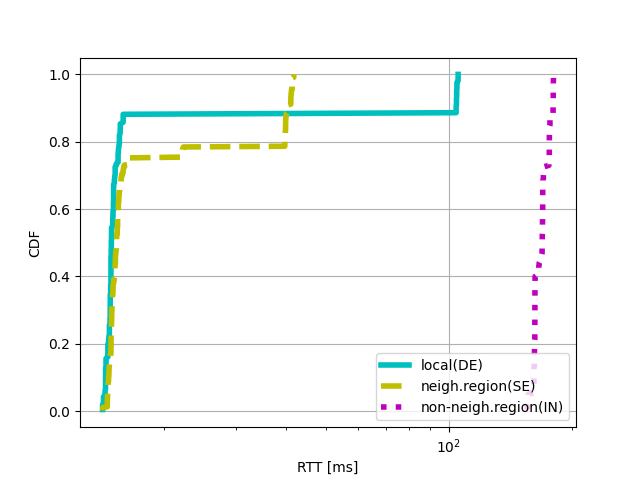}}
	\caption{Latency to assigned replica servers (/24) for Akamai, Amazon CloudFront and Fastly for resources of top50 German websites. The client in Germany uses a recursive DNS resolvers local (cyan) and distant ones in Sweden and India.}
	\label{fig:DNS_localization_DE}
\end{figure}

Figure~\ref{fig:DNS_localization_DE} shows the full set of the graphs to illustrate the importance of DNS for replica selection for the client in Germany using a local DNS resolver (cyan) and distant ones in Sweden and India.

\subsection{Performance and Third-party DNS}

\begin{figure}
	\centering
	\subfigure[AR\label{fig:ARResolution}]{\includegraphics[width=0.48\columnwidth]{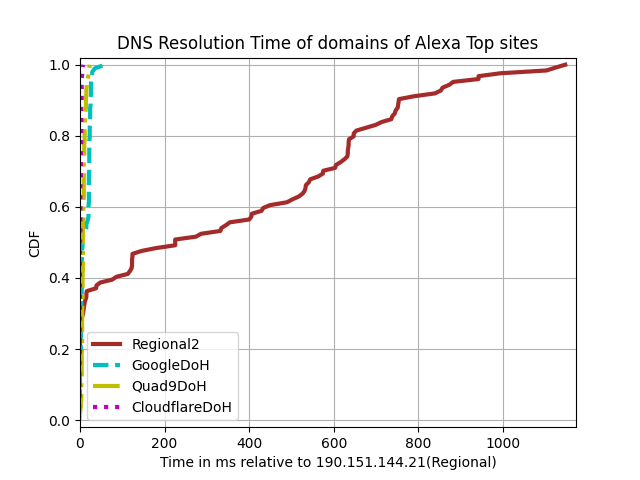}} 
	\subfigure[DE\label{fig:DEResolution}]{\includegraphics[width=0.48\columnwidth]{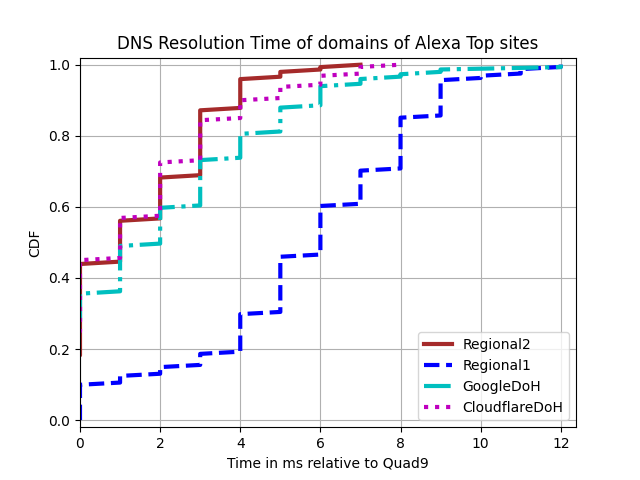}}	
	\caption{DNS resolution times of individual, public DoH and DNS providers in the Argentina and Germany, relative to the best performing DNS service, across each country.}
	\label{fig:resolutiontime_a}
\end{figure}

Figure~\ref{fig:resolutiontime_a} shows the remaining graphs to show how resolution times of different resolvers varies in different locales.

\subsection{Sharding Performance and Third-party DNS}

\begin{figure*}
	\centering
	\subfigure[US\label{fig:USResolution_sharding_a}]{\includegraphics[width=0.48\columnwidth]{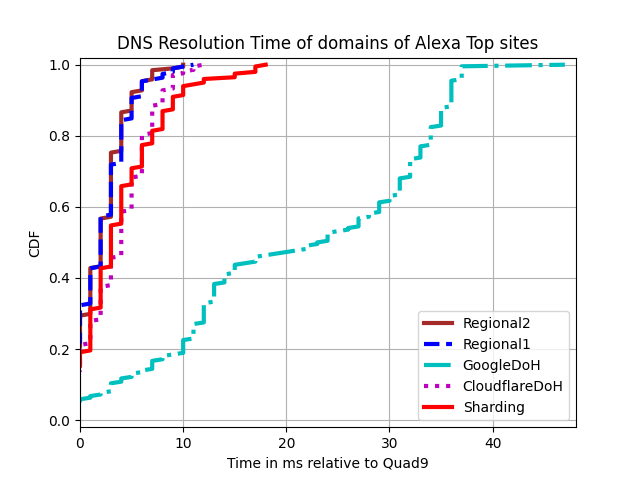}} 
	\subfigure[AR\label{fig:ARResolution_sharding_a}]{\includegraphics[width=0.48\columnwidth]{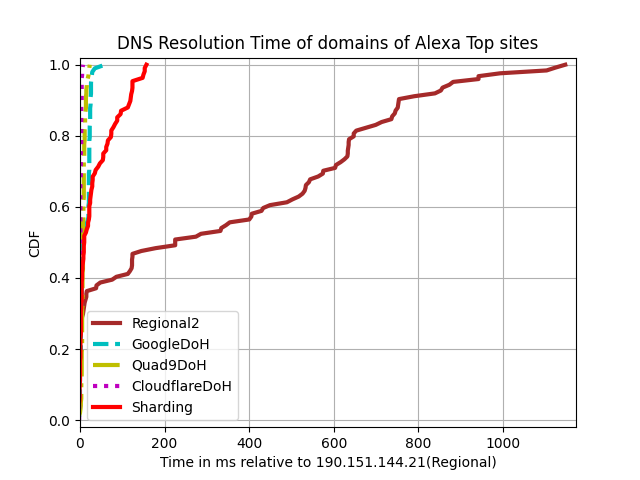}} 
	\subfigure[IN\label{fig:INResolution_sharding_a}]{\includegraphics[width=0.48\columnwidth]{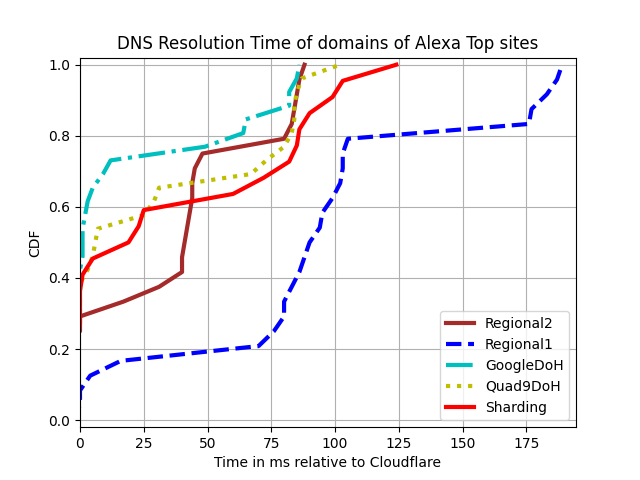}}
	\caption{DNS resolution times of individual, public DoH and DNS providers and sharding in the Argentina, India and the US relative to the best performing DNS service, across each country.}
	\label{fig:resolutiontime_sharding_a}
\end{figure*}

Figure~\ref{fig:resolutiontime_sharding_a} shows the remaining graphs to show that resolution times of different resolvers and sharding varies in different locales.

\subsection{The Impact of CDN and DNS on Web QoE across locales}

\begin{figure}
	\centering
	\includegraphics[width=0.3\textwidth]{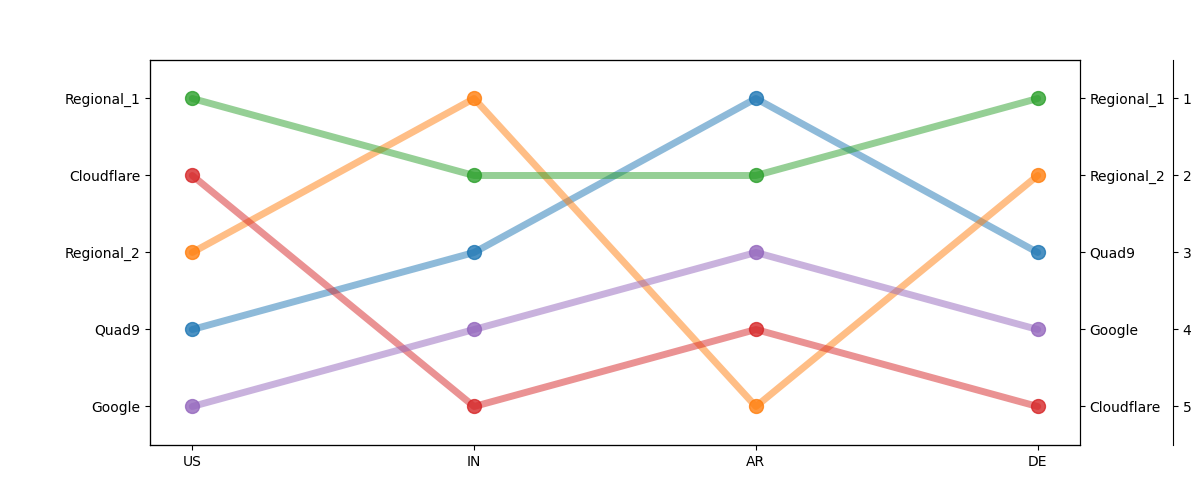}
	\caption{DNS resolvers ranking by Time-to-First-Byte for resources hosted on Fastly across different countries. }
	\label{fig:ttfbRankinga}
\end{figure}

Figure~\ref{fig:ttfbRankinga} shows the graph for the ranking of public and regional DNS resolvers by TTFB for resources hosted on Fastly in different locales.

\begin{figure}
	\centering
	\subfigure[US,Amazon\label{fig:US_OnomavsProxyAmazon}]{\includegraphics[width=0.2\textwidth]{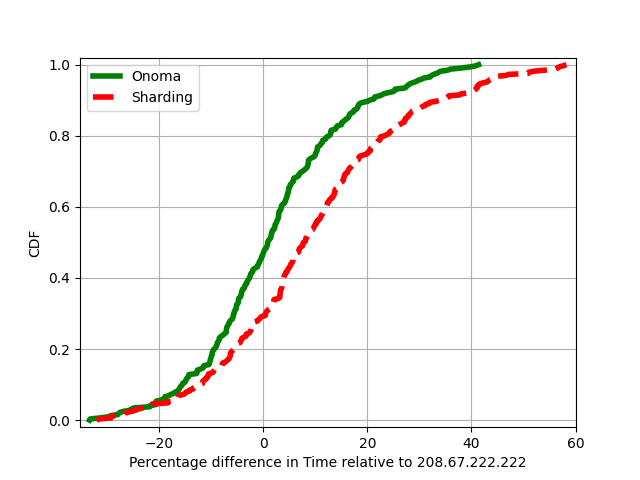}} 
	\subfigure[US,Fastly\label{fig:US_OnomavsProxyFastly}]{\includegraphics[width=0.2\textwidth]{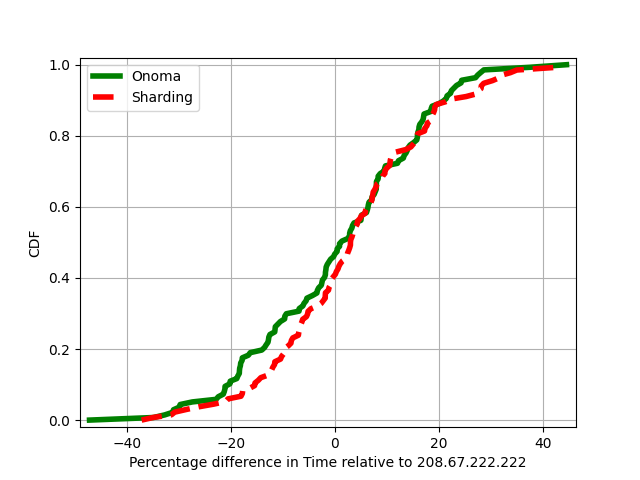}}
	\subfigure[AR,Amazon\label{fig:AR_OnomavsProxyAmazon}]{\includegraphics[width=0.2\textwidth]{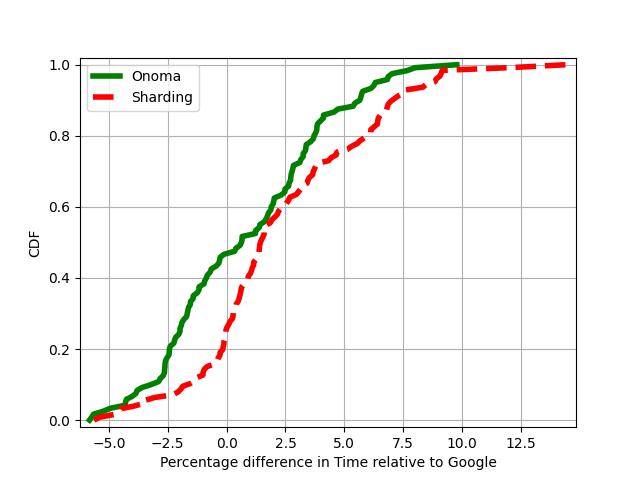}}
	\subfigure[AR,Fastly\label{fig:AR_OnomavsProxyFastly}]{\includegraphics[width=0.2\textwidth]{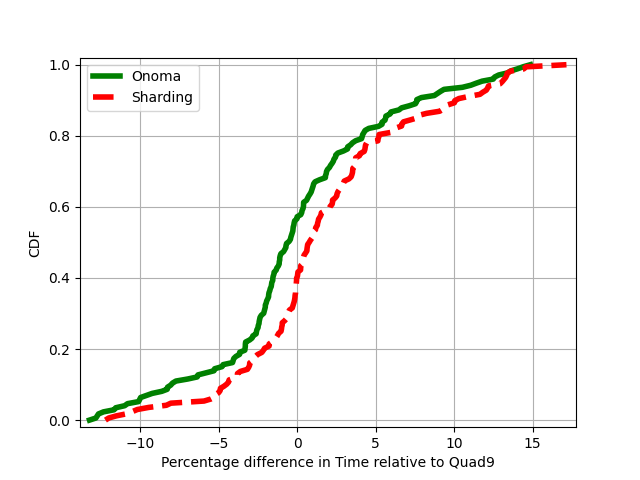}}
	\caption{Percentage Difference in Time-to-First-Byte of Sharding and \name{} relative to the best DNS service across the US and Argentina. For every CDN, country combination shown in the graph \name{} performs better than sharding.}
	\label{fig:OnomavsProxy}
\end{figure}

\subsection{End-to-end Evaluation of \name{}}

\begin{figure}
	\centering
	\subfigure[AR,Amazon\label{fig:AR_amazon_ttfb}]{\includegraphics[width=0.2\textwidth]{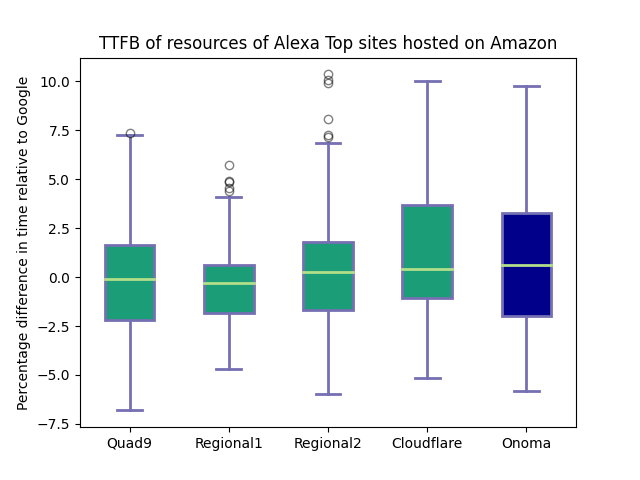}}
	\subfigure[AR,Fastly\label{fig:AR_fastly_ttfb}]{\includegraphics[width=0.2\textwidth]{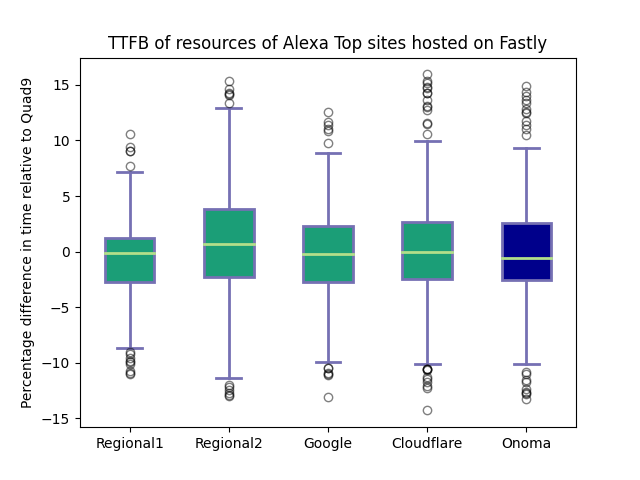}}
	\subfigure[IN,Amazon\label{fig:IN_amazon_ttfb}]{\includegraphics[width=0.2\textwidth]{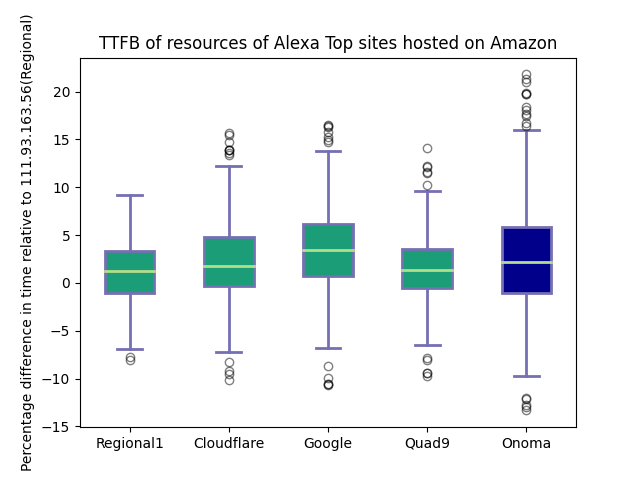}}
	\subfigure[IN,Fastly\label{fig:IN_fastly_ttfb}]{\includegraphics[width=0.2\textwidth]{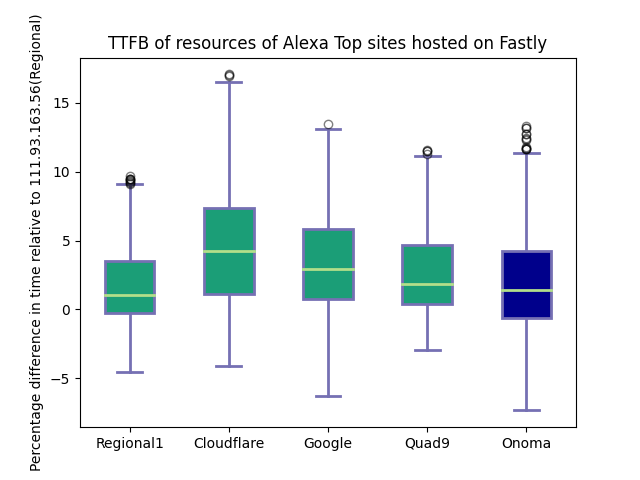}}
	\subfigure[US,Amazon\label{fig:US_amazon_ttfb}]{\includegraphics[width=0.2\textwidth]{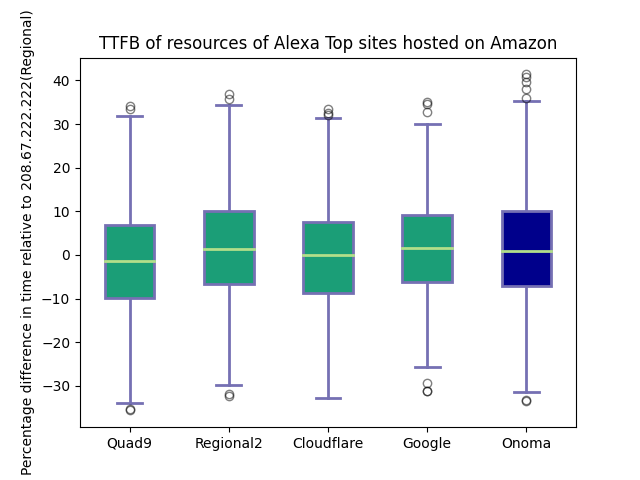}} 
	\subfigure[US,Fastly\label{fig:US_fastly_ttfb}]{\includegraphics[width=0.2\textwidth]{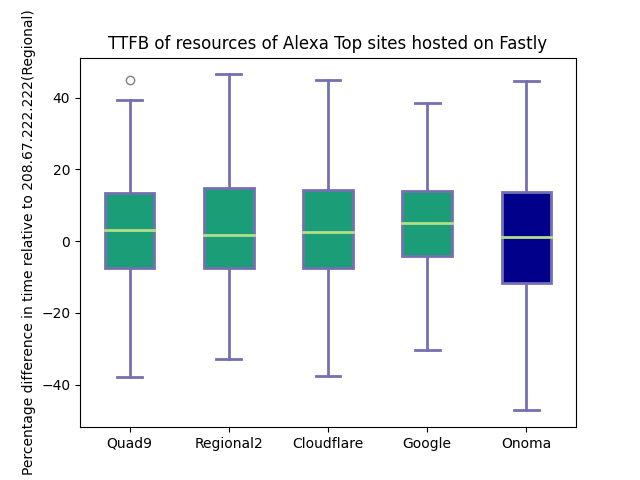}}
	\caption{Performance of individual, public DoH and DNS resolvers in each country and using \name{}, relative to the best DNS service across different CDNs and countries.}
	\label{fig:ttfb_wOnomaappendix}
\end{figure}

Figure~\ref{fig:ttfb_wOnomaappendix} shows the remaining graphs for the end-to-end evaluation of \name{}, comparing \name{} to other DNS providers for content hosted on different CDNs across the locales.

\end{document}